\documentclass[useAMS,usenatbib]{mn2e}
\usepackage{graphicx}

\title[Electromotive force in the BZ process]{Electromotive force in the Blandford-Znajek process}
\author[K. Toma and F. Takahara]{Kenji Toma$^{1,2,3}$\thanks{E-mail:toma@astr.tohoku.ac.jp} 
and Fumio Takahara$^{3}$\\
$^{1}$Astronomical Institute, Tohoku University, Sendai 980-8578, Japan\\
$^{2}$Frontier Research Institute for Interdisciplinary Sciences, Tohoku University, Sendai 980-8578, Japan\\
$^{3}$Department of Earth and Space Science, Graduate School of Science, Osaka University, Toyonaka 560-0043, Japan}

\begin{document}

\date{}

\pagerange{\pageref{firstpage}--\pageref{lastpage}} \pubyear{}

\maketitle

\label{firstpage}

\begin{abstract}
One of the mechanisms widely considered for driving relativistic jets in active galactic nuclei,
galactic microquasars, and gamma-ray bursts is 
the electromagnetic extraction of the rotational energy of a central black hole,
i.e., the Blandford-Znajek process, although the origin of the electromotive force 
in this process is still under debate. 
We study this process as the steady unipolar induction in the Kerr black hole magnetosphere
filled with a collisionless plasma screening the electric field (the $\bmath{D}$ field) along the 
magnetic field (the $\bmath{B}$ field), i.e., $\bmath{D}\cdot \bmath{B}$ = 0.
We extend the formulations and arguments 
made by Komissarov, and generally show that the origin of the electromotive 
force is ascribed to the ergosphere. It is explicitly shown
that open magnetic field lines penetrating the ergosphere have a region
where the $\bmath{D}$ field is stronger than the $\bmath{B}$ field
in the ergosphere, and it keeps driving the poloidal 
currents and generating the electromotive force and the outward Poynting flux. 
The range of the possible value of the so-called angular velocity of the magnetic field line
$\Omega_{\rm F}$ is deduced for the field lines threading the equatorial plane in the ergosphere.
We briefly discuss the relation between our conclusion and the ideal magnetohydrodynamic 
condition.
\end{abstract}


\section{Introduction}
\label{sec:intro}

Collimated outflows, or jets, with relativistic speeds are observed in active galactic
nuclei and galactic microquasars, and are presumably driven in gamma-ray bursts.
Possible mechanisms for driving relativistic jets include the electromagnetic extraction of 
the rotational energy of a central black hole (BH) \citep{blandford77}, the electromagnetic
extraction of the rotational energy of an accretion flow around a BH \citep{lovelace76}, 
and the thermal gas ejection from an accretion flow 
\citep[e.g.,][]{paczynski90,asano09,becker11,toma12}.

The electromagnetic extraction of the rotational energy of the accretion flow is the same 
mechanism as pulsar winds. \cite{goldreich69} showed that when the poloidal magnetic field is 
penetrating a rotating conductive star, the steady force-free magnetosphere can be established 
outside the star, where the poloidal electric fields are maintained by the charge distribution
and the poloidal currents flow (i.e., the unipolar induction process).
As a result, outward electromagnetic energy and angular momentum fluxes continue to be generated.
The origin of the electromotive force (and the fluxes) is the matter-dominated rotating star,
while the plasma outside the star only supports the electromagnetic field, playing a passive role
for keeping the electric potential differences in the whole system.

As for the energy extraction of the BH itself, \cite{blandford77} 
first showed a mathematical solution for the electromagnetic energy and angular momentum fluxes 
extracted from the BH with the force-free approximation in the slow rotation limit of the BH, 
in a similar mechanism to the Goldreich-Julian model.
It has been frequently argued that this electromagnetic process may be the only viable process to 
extract the rotational energy from a BH \citep[e.g.,][]{bejger12,bardeen72,penrose69}.
However, the origin of the electromotive force in this process is still being debated.
There is no matter-dominated region in which the poloidal magnetic field is anchored, and
all the matter in the BH magnetosphere only plays a passive role.
Then it is not a simple question what determines the electric potential difference 
(or the so-called angular velocity of the magnetic field line $\Omega_{\rm F}$) 
and drives the poloidal currents. The literature on the origin of 
the electromotive force focuses on either the event horizon \citep{thorne86}, 
the wind separation surface \citep{takahashi90,beskin00,levinson06,beskin10,okamoto12}, or the ergosphere
\citep{komissarov04,komissarov09} \citep[see also][]{menon05,menon11,contopoulos13}, although
the first one is unlikely because the event horizon is causally
disconnected from the BH exterior \citep{punsly89,punsly08,beskin00}. 

The above literature mainly uses the analytical methods, while the numerical simulations based on 
the ideal magnetohydrodynamic (MHD) approximation have been actively performed to investigate the nature of
the BH magnetosphere (and even its interaction with the accretion flow) 
\citep[e.g.,][]{koide02,komissarov05,mckinney06,barkov09,tchekho11}.
Based on the simulation results, it was proposed
that the unavoidable rotation of the fluids in the ergosphere generates MHD waves propagating
outwards, and its feedback causes the fluids to have negative hydrodynamic energy as measured at infinity.
They plunge into the BH, decreasing its rotational energy \citep{koide02}. 
However, the long-term simulations have demonstrated that no regions of negative hydrodynamic energy are 
seen in the steady state \citep{komissarov05}, and thus more studies are 
required to understand the real mechanism of the extraction of the BH rotational energy 
\citep{komissarov09}. 
Furthermore, it has not been discussed what determines the value of $\Omega_{\rm F}$ and drives 
the poloidal currents by utilizing the MHD simulation results, as far as we are aware.

Since the Blandford-Znajek process is purely electromagnetic, the force-free simulations appear
more suitable than the MHD simulations for investigating the nature of its mechanism (without  
including the complicated interaction of the BH magnetosphere with the accretion flow). 
\cite{komissarov04} clearly shows that when the BH exterior is vacuum, the electric
field (the $\bmath{D}$ field) can be stronger than the magnetic field 
(the $\bmath{B}$ field) in the ergosphere, and 
performed the resistive force-free numerical simulations of the plasma-filled BH magnetosphere, 
demonstrating that such a strong electric field drives the poloidal currents.
A similar conclusion was drawn from simulation results of the force-free electromagnetic field on 
highly compact regular space-times (i.e., with no event horizon) with an ergosphere \citep{ruiz12}. 
However, it has not been evident why the electric field is maintained to be strong in the 
plasma-filled BH magnetosphere and how the steady state with $\Omega_{\rm F} > 0$ and non-zero poloidal
currents is realized. A related question is
whether the steady state with $\Omega_{\rm F} = 0$ and/or zero poloidal currents is possible or not.

In this paper, we extend the argument of \cite{komissarov04}, providing more general insights.
We consider the steady, axisymmetric
Kerr BH magnetosphere filled with a collisionless plasma satisfying 
$\bmath{D} \cdot \bmath{B} = 0$, and
show in a fully analytical way that for open magnetic field lines threading the ergosphere,
the state with $\Omega_{\rm F} > 0$ and non-zero poloidal currents is forced to be maintained, i.e.,
the outward Poynting flux inevitably keeps being generated. The origin of the electromotive 
force is ascribed to the ergosphere. 
We show a self-consistent electromagnetic structure along an open field line threading the 
equatorial plane in the ergosphere, and deduce the range of the possible value of $\Omega_{\rm F}$ for such field lines.\footnote{
\cite{punsly90} proposed a MHD model for the unipolar induction for the poloidal magnetic field
threading the equatorial plane in the ergosphere, but they argue that the potential difference is determined
by the particle inertia in a similar way to the pulsar wind case. Such a situation is different from
that we consider in this paper.
} 
Our assumptions for the magnetospheric plasma are stated in Section~\ref{sec:BHmag1},
which do not rely on a specific Ohm's law as done in \cite{komissarov04}, 
the force-free condition ($F_{\mu\nu} I^\nu = 0$), nor the ideal MHD condition 
($F_{\mu\nu} u^\nu = 0$).
 
This paper is organized as follows. We first review the Goldreich-Julian model of pulsar winds in
Section~\ref{sec:pulsar} and the convenient formulation of the electrodynamics in Kerr space-time
in Section~\ref{sec:BHEM}. We then study the plasma-filled BH magnetosphere, discussing its 
general properties (Section~\ref{sec:BHmag}) and 
the origin of the electromotive force (Section~\ref{sec:BHuni}).
Section~\ref{sec:discussion} is devoted to summary and discussion.

\section{Unipolar Induction of Pulsars}
\label{sec:pulsar}

We review the unipolar induction mechanism of pulsars, first discussed by \cite{goldreich69},
{\it which is quite helpful for understanding the unipolar induction of rotating BHs.}
In particular, it is instructive that the electric field outside the star is stronger 
in the plasma-filled case than that in the vacuum case (equations~\ref{eq:pulsarE1} and \ref{eq:pulsarE2})
and that the star drives the poloidal currents inside itself in the direction of $-\bmath{E}$,
which generates the Poynting flux (equation~\ref{eq:pulsar_poynting}). We will refer to these points
in Section~\ref{sec:active} and in Section~\ref{sec:structure}, respectively.

\subsection{Electric Potential Differences}
\label{sec:pulsar_potential}

Let us consider a uniformly magnetized, rotating star with angular velocity $\Omega_{\rm s}$.
Our assumptions are as follows: (1) The rotation vector is parallel to the magnetic dipole
moment, so that the system is steady and axisymmetric. (2) At and inside the stellar surface,
the matter is highly conductive, and its rotational energy dominates the total energy.
(3) The region outside the star is filled with plasma. The plasma is dilute and
collisionless, but its number density is high enough to screen the electric field along the magnetic
field lines, i.e., $\bmath{E} \cdot \bmath{B} =0$ is satisfied. 
The energy density of the particles is much smaller than that of the electromagnetic field.
The magnetic field is so strong that the gravitational force of 
the central star is negligible compared with the Lorentz force. In this section, we adopt
the units of $c=1$.

First of all, at and inside the stellar surface, the high conductivity implies 
\begin{equation}
\bmath{E} = - \bmath{V} \times \bmath{B} = -\varpi \Omega_{\rm s} \bmath{e}_\varphi \times \bmath{B},
\label{eq:Estar}
\end{equation}
where $\bmath{V}$ is the fluid velocity, and we have adopted the cylindrical coordinate system
$(\varpi, \varphi, z)$. This electric field is produced by the non-zero charge density
$\rho = (1/4\pi)\nabla \cdot \bmath{E}$.

The electromagnetic field structure in the region outside the star, i.e., magnetosphere,
is described as follows. In the axisymmetric configuration, the poloidal magnetic field
can be written as
\begin{equation}
\bmath{B}_p = \frac{1}{\varpi} \nabla \Psi \times \bmath{e}_\varphi,
\end{equation}
where $\Psi(\varpi, z)$ is called the flux function, representing the toroidal component of the
vector potential times $\varpi$, or the magnetic flux threading through a circle of 
$\varpi$ for a given $z$. The above equation leads to $\bmath{B}\cdot \nabla \Psi=0$,
which means that $\Psi(\varpi, z)$ is constant along each magnetic field line.
As for the electric field, Faraday's law in the steady state gives $\nabla \times \bmath{E} =0$.
The axisymmetric condition implies $E_\varphi = 0$. This equation and 
$\bmath{E} \cdot \bmath{B} = 0$ mean that the electric field can be written as
\begin{equation}
\bmath{E} = -\varpi \Omega_{\rm F} \bmath{e}_\varphi \times \bmath{B},
\label{eq:E}
\end{equation}
where $\Omega_{\rm F}$ is a function of $(\varpi, z)$. Substituting this equation into
$\nabla \times \bmath{E} = 0$ and using $\nabla \cdot \bmath{B} = 0$, one obtains
\begin{equation}
\bmath{B} \cdot \nabla \Omega_{\rm F} = 0.
\end{equation}
This means that $\Omega_{\rm F}$ is constant along each magnetic field line. The electric
field is also described by $\bmath{E} = -\Omega_{\rm F} \nabla\Psi$, which means that each
magnetic field line is equi-potential, and $\Omega_{\rm F}$ corresponds to the potential
difference between the field lines. The plasma in the magnetosphere sustains 
$\rho = (1/4\pi)\nabla \cdot \bmath{E}$, which produces the electric field
in the similar manner to the star.

Equation~(\ref{eq:E}) is satisfied in the whole system we now consider including 
the inside of the star, as long as there is no region where 
$\bmath{E} \cdot \bmath{B} \neq 0$, i.e., no gap. Comparing it with equation~(\ref{eq:Estar}),
one obtains
\begin{equation}
\Omega_{\rm F} = \Omega_{\rm s}.
\label{eq:F*}
\end{equation}
This means that the stellar rotation generates the potential differences in the 
magnetosphere, i.e., the origin of the electromotive force for pulsar winds is ascribed to
the rotation of the stellar matter.
For the rigid rotation of the star, $\Omega_{\rm F}$ is constant in the whole system.

It is interesting that the plasma enhances the strength of the electric field across the magnetic field,
although it screens that along the magnetic field (satisfying $\bmath{E} \cdot \bmath{B} = 0$).
In fact, in the plasma-filled case for the dipole
magnetic field, which we have considered, one has from equation~(\ref{eq:E})
\begin{equation}
E_r = \mu \Omega_{\rm F} \frac{\sin^2 \theta}{r^2}, ~~ 
E_{\theta} = - \mu \Omega_{\rm F} \frac{2 \sin\theta \cos\theta}{r^2},
\label{eq:pulsarE1}
\end{equation}
in terms of the spherical coordinates $(r, \theta, \varphi)$,
where $\mu$ is the magnetic dipole moment. In the vacuum case, the electric field has quadrupole
shape \citep{goldreich69}
\begin{equation}
E_r = - \mu \Omega_{\rm F} r_{\rm s}^2 \frac{3\cos^2\theta -1}{r^4}, ~~
E_{\theta} = - \mu \Omega_{\rm F} r_{\rm s}^2 \frac{2 \sin\theta \cos\theta}{r^4},
\label{eq:pulsarE2}
\end{equation}
where $r_{\rm s}$ is the radius of the star.
It is obvious that the electric field is stronger in the plasma-filled case for $r > r_{\rm s}$. 

Let us consider the particle motions in the magnetosphere. Since the energy density is dominated
by the electromagnetic field, the particles move via the $\bmath{E} \times \bmath{B}$ drifts.
For the dipole magnetic field and the associated electric field shown above, this motion is in the
direction of $\varphi$. It carries the drift currents, 
but does not produce $B_\varphi$. The drift
velocity is written by using equation~(\ref{eq:E}) as
\begin{equation}
\mathbf{v}_d = \varpi\Omega_{\rm F} \bmath{e}_{\varphi} 
- \varpi\Omega_{\rm F} \frac{B_{\varphi}}{B^2} \bmath{B}.
\label{eq:v_d}
\end{equation}
As long as $B_{\varphi} = 0$, one has $\bmath{v}_d = \varpi \Omega_{\rm F} \bmath{e}_\varphi$.
Because $\Omega_{\rm F} = \Omega_{\rm s}$, the particles co-rotate with the star.
The drift current is $J_\varphi = \rho \varpi \Omega_{\rm s}$.
Now one can calculate the charge density in the magnetosphere at $\varpi < 1/\Omega_{\rm F}$ as
\begin{equation}
\rho = \frac{1}{4\pi} \nabla \cdot \mathbf{E} = \frac{- \Omega_{\rm s} B_z}{2\pi 
(1-\varpi^2 \Omega_{\rm s}^2)} \approx \frac{-\Omega_{\rm s} B_z}{2\pi},
\label{eq:rho_GJ}
\end{equation}
This implies that $\rho <0$ ($>0$) in the region where $B_z >0$ ($<0$).

\subsection{Poloidal Current Structure}
\label{sec:pulsar_current}

The co-rotation of the particles cannot be maintained outside the light cylinder at 
$\varpi = \varpi_{\rm lc} \equiv 1/\Omega_{\rm F}$. The poloidal electric currents must flow. 
The poloidal currents produce
$B_{\varphi}$, which prevents the particle speeds from exceeding the light speed (see equation~\ref{eq:v_d}).
The particles must flow out along the magnetic field lines which cross the light cylinder. Those 
magnetic field lines must be open, because of the symmetry with respect to the equatorial plane.
The charge density distribution (equation~\ref{eq:rho_GJ}) and the particle outflow imply that the
poloidal electric currents flow towards the northern and southern poles of the star and flow 
out from the light cylinder around the equatorial plane. Therefore one finds $B_\varphi < 0$ ($>0$) in the main body
of the outflow region (i.e., the open field line region) in the northern (southern) hemisphere.

According to equation~(\ref{eq:v_d}), one has $v_{d,\varphi} = (1-B_\varphi^2/B^2) \varpi \Omega_F$, and 
thus it is reasonable that $|B_\varphi| \sim |\bmath{B}_p|$ around the light cylinder.
One might consider that such a strong $B_\varphi$ is produced by the poloidal currents flowing
due to the effects of the particle inertia around the light cylinder,
but this is not the case, because the particle energy density is still much smaller
than the electromagnetic field energy density around the light cylinder, 
as seen by the MHD theory \citep[cf.,][]{beskin10,toma13}.
$B_\varphi$ must be produced by the poloidal currents which are driven by the
matter-dominated star. The outflows
of the positive (negative) charges from the high (low) potential region in the star result in 
the net electric field $\bmath{E} + \bmath{V} \times \bmath{B} \neq 0$ in the direction of $-\bmath{E}$,
which continues driving the poloidal currents and supplying the positive (negative) charges to 
the high (low) potential region.

The currents flowing in the direction of $-\bmath{E}$ in the star generates the poloidal component of
the Poynting flux $\bmath{S}_p = \bmath{E} \times \bmath{B}_{\varphi}/4\pi 
= -\Omega_F \varpi B_\varphi \bmath{B}_p/4\pi$, as described by
\begin{eqnarray}
\nabla \cdot \bmath{S}_p &=& -\bmath{E} \cdot \bmath{J}_p
= (\bmath{V}_{\varphi} \times \bmath{B}_p) \cdot \bmath{J}_p \nonumber \\
&=& -(\bmath{J}_p \times \bmath{B}_p) \cdot \bmath{V}_{\varphi}.
\label{eq:pulsar_poynting}
\end{eqnarray}
This equation also means that the Lorentz force $\bmath{J}_p \times \bmath{B}_p$, being in the direction
of $-\bmath{V}_\varphi$, decelerates the stellar rotation as a feedback of the Poynting flux generation.
In this way, the rotating star keeps generating the electromotive force (equation~\ref{eq:F*}), driving the
poloidal electric currents, and generating the outward Poynting flux.

Since $|B_\varphi| \sim |\bmath{B}_p|$ is expected around the light cylinder, one may estimate
$|\bmath{S}_p| \sim B_{p,{\rm lc}}^2/4\pi$, where $B_{p,{\rm lc}}$ is the strength of the poloidal magnetic field
at the light cylinder. The dipole field implies 
$B_{p,{\rm lc}} \sim B_{p,{\rm s}} (\varpi_{\rm lc}/r_{\rm s})^{-3}$, where $B_{p,{\rm s}}$ is the 
strength of the poloidal magnetic 
field at the stellar surface. Then the electromagnetic luminosity is roughly estimated as
\begin{equation}
L \sim 4\pi r_{\rm lc}^2 |\bmath{S}_p| \sim B_{p,{\rm s}}^2 \Omega_{\rm s}^4 r_{\rm s}^6.
\label{eq:luminosity}
\end{equation}
Here the condition $|B_\varphi| \sim |\bmath{B}_p|$ around the light cylinder is equivalent to
the estimate of the total current $I$ flowing towards the star. In fact, the order of the current density is 
probably $J_p \sim |\rho| c$, and then $I \sim J_p A$, where $A$ is the area of the stellar surface into which
the currents flow, and estimated to be $A \simeq \pi r_{\rm s}^3 \Omega_{\rm s}$
\citep{goldreich69}. The electromagnetic luminosity is also calculated by $L \sim 2 I \Phi$,
where the potential difference $\Phi \sim B_{p, {\rm s}} r_{\rm s}^3 \Omega_{\rm s}^2$, which is
consistent with equation~(\ref{eq:luminosity}). (The factor 2 comes from the contributions from the 
northern and southern hemispheres.)

The poloidal currents must cross the 
poloidal magnetic field lines in the direction of $\bmath{E}$ outside the star for making
closed circuits, which accelerates the particles via the $\bmath{J}_p \times \bmath{B}_\varphi$
force and then converts the Poynting flux to the particle kinetic energy flux. In this paper
we do not discuss the particle acceleration mechanism in detail 
\citep[see][and references therein]{beskin10,toma13}.

\section{Electrodynamics in Kerr Space-time}
\label{sec:BHEM}

We briefly summarize the geometrical properties of Kerr space-time and the electrodynamics
in order to define the various physical quantities used in this paper. We follow
the definitions and formulations of \citet{komissarov04} except for keeping
$4\pi$ in Maxwell equations. 
We adopt the units of $c = 1$ and $GM=1$, where $M$ is the BH mass. 
For such units, the gravitational radius $r_g = GM/c^2 =1$.

\subsection{The 3+1 Decomposition of Space-time}
 
The space-time metric can be generally written as
\begin{equation}
ds^2 = g_{\mu \nu} dx^\mu dx^\nu = 
-\alpha^2 dt^2 + \gamma_{ij} (\beta^i dt + dx^i) (\beta^j dt + dx^j),
\end{equation}
where $\alpha$ is called the lapse function, $\beta^i$ the shift vector, and $\gamma_{ij}$ the three-dimensional
metric tensor of the space-like hypersurfaces. The hypersurfaces are regarded as the absolute 
space at different instants of time $t$ \citep[cf.][]{thorne86}. For Kerr space-time, 
$\partial_t g_{\mu \nu} = \partial_\varphi g_{\mu \nu} = 0$. These correspond 
to the existences of the Killing vector fields $\xi^\mu$ and $\chi^\mu$. 
In the coordinates $(t, \varphi, r, \theta)$, $\xi^\mu = (1, 0, 0, 0)$ and 
$\chi^\mu = (0, 1, 0, 0)$.

The local fiducial observer \citep[FIDO;][]{bardeen72,thorne86}, whose world line is 
perpendicular to the absolute space, is described by the coordinate four-velocity
\begin{equation}
n^\mu = \left(\frac{1}{\alpha}, \frac{-\beta^i}{\alpha} \right), ~~~
n_\mu = g_{\mu \nu} n^{\nu} = (-\alpha, 0, 0, 0).
\end{equation}
The angular momentum of this observer is 
$n \cdot \chi = g_{\mu \nu} n^\mu \chi^\nu = n_\varphi = 0$, and thus
FIDO is also a zero angular momentum observer \citep[ZAMO;][]{thorne86}. 
Note that the FIDO frame is not inertial, but it can be used as a convenient orthonormal basis 
to investigate the local physics \citep{thorne86,punsly90,punsly08}. 

\begin{figure}
\begin{center}
\includegraphics[scale=0.7]{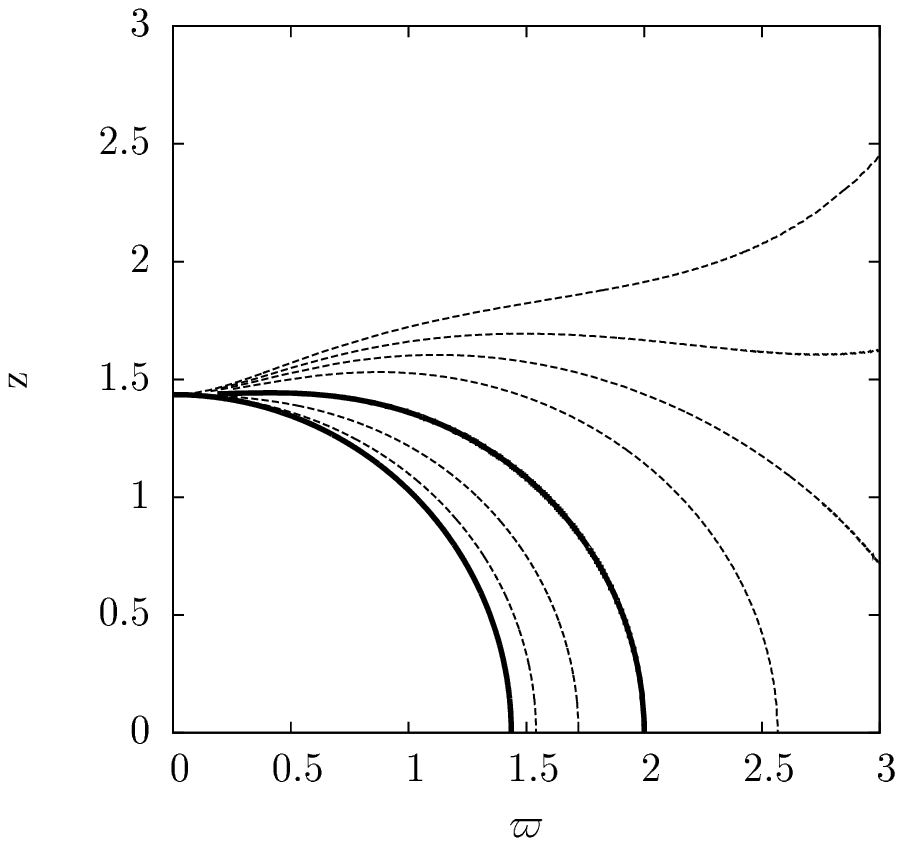}
\end{center}
\caption{
The event horizon (inner thick line) and the outer boundary of the ergosphere (outer thick line)
of Kerr space-time. The thin lines represent
$\Omega - \alpha/\sqrt{\gamma_{\varphi \varphi}} = 0.2, 0.1, -0.1, -0.14, -0.17, -0.2$ in the BL 
coordinates in the order of increasing $r$. The line of 
$\Omega - \alpha/\sqrt{\gamma_{\varphi\varphi}} = 0$ is identical to the outer thick line.
The spin parameter is set to be $a=0.9$.
}
\label{fig:fun}
\end{figure}

In the Boyer-Lindquist (BL) coordinates $(t, \varphi, r, \theta)$ (see Appendix~\ref{sec:app1}),
FIDOs rotate with the coordinate angular velocity
\begin{equation}
\Omega \equiv \frac{d\varphi_{\rm FIDO}}{dt} = -\beta^{\varphi} >0, 
\end{equation}
which is in the same direction as the BH.
The BL coordinates have the well-known coordinate singularity ($g_{rr} = \infty$) at the
event horizon. The radius of the event horizon
is $r_{\rm H} = 1+ \sqrt{1-a^2}$. The Killing vector $\xi^\mu$
is space-like in the ergosphere, where $\xi^2 = g_{tt} = - \alpha^2 + \beta^2 > 0$.
The radius of the outer
boundary of the ergosphere (i.e., the stationary limit) is $r_{\rm es} = 1 + \sqrt{1-a^2 \cos^2\theta}$.
At infinity, this space-time asymptotes to the flat one. The shapes of the event horizon and the
ergosphere are shown in Fig.~\ref{fig:fun}.

The Kerr-Schild (KS) coordinates have no coordinate singularity at the event horizon.
However, the KS spatial coordinates are no longer orthogonal ($\gamma_{r\varphi} \neq 0$; see 
Appendix~\ref{sec:app1}), and then one should be cautious for examining the spatial structure of the
electromagnetic field by using the KS coordinates.

\subsection{The 3+1 Electrodynamics}

In order to study the test electromagnetic field in Kerr space-time, we adopt the 3+1 electrodynamics
of the version which was developed by \cite{komissarov04} 
\citep[see also][and references therein]{landau75,komissarov09}.\footnote{
\citet{thorne82} and \citet{thorne86} developed the 3+1 electrodynamics of the version without introducing
$\bmath{E}$ or $\bmath{H}$, and showed some of the expressions in this paper, such as equations 
(\ref{eq:Psi}) and (\ref{eq:D}).
}
The covariant Maxwell equations $\nabla_\nu {}^*F^{\mu\nu} = 0$ and $\nabla_\nu F^{\mu\nu} = 4\pi I^\mu$ are
reduced to
\begin{equation}
\nabla \cdot \bmath{B} = 0, ~~~~ \partial_t \bmath{B} + \nabla \times \bmath{E} = 0, 
\label{eq:maxwell1}
\end{equation}
\begin{equation}
\nabla \cdot \bmath{D} = 4\pi \rho, ~~~~ - \partial_t \bmath{D} + \nabla \times \bmath{H} = 4\pi \bmath{J},
\label{eq:maxwell2}
\end{equation}
where $\nabla \cdot \bmath{C}$ and $\nabla \times \bmath{C}$ denote 
$(1/\sqrt{\gamma})\partial_i(\sqrt{\gamma} C^i)$ and $e^{ijk} \partial_j C_k$, respectively, and 
$e^{ijk} = (1/\sqrt{\gamma}) \epsilon^{ijk}$ is the Levi-Civita pseudo-tensor of the absolute space.
The condition of zero electric and magnetic susceptibilities for general fully-ionized plasmas leads to
following constitutive equations,
\begin{equation}
\bmath{E} = \alpha \mathbf{D} + \bmath{\beta} \times \bmath{B},
\label{eq:relation_E}
\end{equation}
\begin{equation}
\bmath{H} = \alpha \mathbf{B} - \bmath{\beta} \times \bmath{D},
\label{eq:relation_H}
\end{equation}
where $\bmath{C} \times \bmath{F}$ denotes $e^{ijk} C_j F_k$. At infinity, $\alpha=1$ and 
$\bmath{\beta} = 0$, so that $\bmath{E} = \bmath{D}$ and $\bmath{H} = \bmath{B}$. Here $\bmath{D}$,
$\bmath{B}$, and $\rho$ are the electric field, magnetic field, and charge density as measured
by FIDOs, respectively (see Appendix~\ref{sec:app1} for more details).
The current $\bmath{J}$ is related to the current as measured by FIDOs, $\bmath{j}$, as
\begin{equation}
\bmath{J} = \alpha \bmath{j} - \rho \bmath{\beta}.
\label{eq:Jj}
\end{equation}

The covariant energy-momentum equation of the electromagnetic field
$\nabla_\nu T^{\nu}_{\mu} = -F_{\mu\nu} I^\nu$ gives us the energy equation as
\begin{equation}
\partial_t \left[\frac{1}{8\pi} (\bmath{E} \cdot \bmath{D} + \bmath{B} \cdot \bmath{H}) \right] + 
\nabla \cdot \left(\frac{1}{4\pi}\bmath{E} \times \bmath{H} \right)= - \bmath{E} \cdot \bmath{J},
\label{eq:energy}
\end{equation}
where $\bmath{C} \cdot \bmath{F}$ denotes $C^i F_i$, and the angular momentum equation as
\begin{eqnarray}
\partial_t \left[\frac{1}{4\pi}(\bmath{D}\times \bmath{B})\cdot \bmath{m}\right] 
+ \nabla \cdot \frac{1}{4\pi} \Bigl[-(\bmath{E}\cdot \bmath{m}) \bmath{D} - 
(\bmath{H}\cdot \bmath{m}) \bmath{B} \nonumber \\
+ \frac{1}{2} (\bmath{E}\cdot \bmath{D} + \bmath{B}\cdot \bmath{H})\bmath{m} \Bigr] 
= -(\rho \bmath{E} + \bmath{J}\times \bmath{B})\cdot \bmath{m},
\label{eq:AM}
\end{eqnarray}
where $\bmath{m} = \partial_\varphi$. From these equations, one can find the energy density,
energy flux, angular momentum density, and angular momentum flux.

\subsection{Steady Axisymmetric Electromagnetic Field in the Vacuum}
\label{sec:wald}

Before investigating the plasma-filled magnetosphere in Kerr space-time, 
the properties of the electromagnetic field in the vacuum (i.e., no plasma) are summarized. 
\cite{wald74} derived the solution of a steady, axisymmetric, vacuum test electromagnetic field 
in Kerr space-time for which 
the magnetic field is uniform, parallel to the rotation axis, at infinity 
in an elegant way by utilizing the fact that $\xi^{\mu}$ and $\chi^{\mu}$ generate
a solution of Maxwell equations (see Appendix~\ref{sec:app1}).
For this solution, the poloidal $\bmath{D}$ field is non-zero.
This may be understood by considering $\nabla \times \bmath{E} = \nabla \times (\alpha \bmath{D}
+ \bmath{\beta} \times \bmath{B}) = 0$. Since 
$\nabla \times (\bmath{\beta} \times \bmath{B}) \neq 0$ generally, one has
$\nabla \times (\alpha \bmath{D}) \neq 0$. This $\bmath{D}$ field is produced by the 
charges at the event horizon and at infinity.
Note that $\bmath{D}\cdot\bmath{B} \neq 0$ for this solution. 

$\nabla \times \bmath{E} = 0$ implies $E_\varphi = 0$, and
$\nabla \times \bmath{H} = 0$ tells us $H_\varphi = 0$.
Then the poloidal components of the energy and angular momentum fluxes are zero. These properties 
are the same as those for the pulsar in the vacuum.

\section{Kerr Black Hole Magnetosphere}
\label{sec:BHmag}

Now we examine the steady, axisymmetric, test electromagnetic field in Kerr space-time 
in which the plasma is filled. As a preparation for the discussion on the unipolar induction
of rotating black holes (in Section~\ref{sec:BHuni}), we summarize the general properties of
the electromagnetic field in Section~\ref{sec:BHmag1}. In Section~\ref{sec:particle}, 
we study the particle motions as viewed by FIDOs. This study provides a conclusion that
$D^2 > B^2$ is the necessary and sufficient condition for driving
the electric currents to flow across the poloidal $\bmath{B}$ field lines.

\subsection{Electromagnetic Field}
\label{sec:BHmag1}

We consider the Kerr BH magnetosphere under the following
assumptions: (1) The poloidal $\bmath{B}$ field produced by the external electric 
currents is penetrating the ergosphere. (2) The plasma in the BH magnetosphere is dilute and 
collisionless, but its number density is high enough to screen the electric field along the
$\bmath{B}$ field lines, i.e.,  $\bmath{D} \cdot \bmath{B} = 0$. The energy density of 
the particles is much smaller than that of the electromagnetic field. 
(3) The gravitational force is negligible compared with the Lorentz force.
(The gravitational force overwhelms the Lorentz force in a region very close to 
the event horizon \citep{punsly08}, but the physical condition in that region hardly affects its exterior.)
These assumptions are the same as those for the 
pulsar case. One big difference is that there is no matter-dominated region on which the $\bmath{B}$
field is anchored (see Section~\ref{sec:intro}). As a result, the determination of $\Omega_{\rm F}$
is not so simple as the pulsar case.

In terms of the vector potential, one can write $\bmath{B} = \nabla \times \bmath{A}$ 
\citep{komissarov04}. Thus one finds
\begin{equation}
B^r = \frac{1}{\sqrt{\gamma}} \partial_\theta \Psi, ~~ B^\theta = \frac{-1}{\sqrt{\gamma}} \partial_r \Psi, 
\label{eq:Psi}
\end{equation}
where we have defined $\Psi \equiv A_\varphi$. It is easily shown that $B^i \partial_i \Psi = 0$,
which means that $\Psi$ is constant along each $\bmath{B}$ field line.

The condition $\bmath{D} \cdot \bmath{B} = 0$ and equation~(\ref{eq:relation_E}) lead to 
$\bmath{E} \cdot \bmath{B} = 0$. Taking account of $E_\varphi = 0$, one can write
\begin{equation}
\bmath{E} = - \bmath{\omega} \times \bmath{B}, ~~~ \bmath{\omega} = \Omega_{\rm F} \bmath{m}.
\label{eq:omegaF_def}
\end{equation}
Substituting this equation into $\nabla \times \bmath{E} = 0$ and $\nabla \cdot \bmath{B} = 0$, 
one obtains
\begin{equation}
B^i \partial_i \Omega_{\rm F} = 0.
\label{eq:omegaF_const}
\end{equation}
That is, $\Omega_{\rm F}$ is constant along each $\bmath{B}$ field line. The $\bmath{E}$ field 
is also described by $E_i = -\Omega_{\rm F} \partial_i \Psi$, which means that each $\bmath{B}$
field line is equi-potential, and $\Omega_{\rm F}$ corresponds to the potential difference 
between the field lines. These properties are the same as those in the pulsar case, 
discussed in Section~\ref{sec:pulsar}. 

In the steady, axisymmetric state, the angular momentum equation (\ref{eq:AM}) is reduced to
\begin{equation}
\nabla \cdot \left(-\frac{H_\varphi}{4\pi} \bmath{B}_p \right) =
B^i \partial_i \left(- \frac{H_\varphi}{4\pi} \right) = - (\bmath{J}_p \times \bmath{B}_p) \cdot \bmath{m}.
\label{eq:AM2}
\end{equation}
The energy equation (\ref{eq:energy}) is reduced to
\begin{equation}
\nabla \cdot \left(-\Omega_F \frac{H_\varphi}{4\pi} \bmath{B}_p \right) =
B^i \partial_i \left(-\Omega_F \frac{H_\varphi}{4\pi} \right) = -\bmath{E} \cdot \bmath{J}_p,
\label{eq:energy2}
\end{equation}
which can be deduced by equation~(\ref{eq:AM2}) together with equations~(\ref{eq:omegaF_def}) and (\ref{eq:omegaF_const}).
These equations imply that $H_\varphi$ is generated by the poloidal currents which have the 
component perpendicular to the poloidal $\bmath{B}$ field, and then the poloidal component of the 
Poynting flux $-\Omega_F H_\varphi \bmath{B}_p / (4\pi)$ is non-zero when $\Omega_{\rm F} \neq 0$.
We note that $\Omega_{\rm F} = -F_{t\theta}/F_{\varphi\theta}$ and $H_\varphi = {}^*F_{t\varphi}$
are the same in the BL and KS coordinates \citep{komissarov04}.

\subsection{Poloidal Currents}
\label{sec:particle}

\citet{komissarov04} argues that the poloidal currents can have the component perpendicular to
the poloidal $\bmath{B}$ field in the region where $D^2 > B^2$, by utilizing a specific Ohm's law (or some
effective resistivity). Here we prove this to be valid by examining particle motions in the 
electromagnetic fields more generally, without relying on the Ohm's law.

We treat the particle motion as viewed by FIDOs. They can use a convenient local orthonormal
basis for which the space-time metric is diagonal and one can investigate local, instantaneous 
particle motions under the Lorentz force as special relativistic dynamics.
In fact, the equation of a particle motion as viewed by FIDOs (either BL or KS FIDOs) is
\begin{equation}
\frac{d \hat{u}_i}{d\hat{t}} = \frac{q}{m}(\hat{D}_i + \epsilon_{ijk} \hat{v}^j \hat{B}^k),
\label{eq:EOM}
\end{equation}
as shown in Appendix~\ref{sec:app2}. 
Here $\bmath{u}$, $\bmath{v}$, $q$, and $m$ are the four-velocity,
three-velocity, charge, and mass of a particle, respectively, and
$\hat{C}_i$ denotes the vector component in respect of the 
FIDO's orthonormal basis, i.e., $\hat{C}_i = C_\mu e^\mu_i$ (see Appendix~\ref{sec:app1}).\footnote{
The quantities $\check{\bmath{E}}$ and $\check{\bmath{B}}$ used in \citet{komissarov04} are 
identical to $\bmath{D}$ and $\bmath{B}$. The FIDO's orthonormal basis is not utilized for discussion
in \citet{komissarov04}.} 
Note that the FIDO frame is not inertial and a particle feels
the gravitational force, although we assume that it is negligible compared with the Lorentz force.
In this equation $\hat{D}_i$ and $\hat{B}^i$ appear as the electric and magnetic fields as viewed by FIDOs, 
respectively, as expected.
The assumption $\bmath{D} \cdot \bmath{B} = 0$ is equivalent to 
$\hat{\bmath{D}} \cdot \hat{\bmath{B}} = 0$, because $D^\mu B_\mu$ is a scalar and 
$\hat{D}^t = D^t = 0$.

When $D^2 < B^2$ is satisfied (which is equivalent to $\hat{D}^2 < \hat{B}^2$), 
the charged particles freely move along the $\hat{\bmath{B}}$ field line and/or
drift in the direction of $\hat{\bmath{D}} \times \hat{\bmath{B}}$.
Effectively one has $\hat{\bmath{u}}_+ \cdot \hat{\bmath{D}} 
=\hat{\bmath{u}}_- \cdot \hat{\bmath{D}} = 0$,
where $\bmath{u}_+$ and $\bmath{u}_-$ denote the four-velocities of positively and negatively
charged particles, respectively.
This equation is valid as long as the plasma is 
dilute and any particle collisions which induce $\hat{\bmath{u}}_+ \cdot \hat{\bmath{D}} \neq 0$ or 
$\hat{\bmath{u}}_- \cdot \hat{\bmath{D}} \neq 0$ are ineffective. 
In the coordinate basis, one has $\bmath{u}_+ \cdot \bmath{D} = \bmath{u}_- \cdot \bmath{D} = 0$,
since $u_\mu D^\mu$ is a scalar and $\hat{D}^t = D^t = 0$.
Then by considering $u^i = u^t v^i$, one has $\bmath{v}_+ \cdot \bmath{D} = \bmath{v}_- \cdot \bmath{D} = 0$.
As a result, the motions of the charged particles can carry only the electric currents satisfying
$\bmath{J}_p \cdot \bmath{D} = 0$ and $\bmath{J}_p \parallel \bmath{B}_p$. 

If $D^2 > B^2$ is realized 
(which is equivalent to $\hat{D}^2 > \hat{B}^2$), then the positively (negatively) charged particles 
are forced to move in the direction of $\hat{\bmath{D}}$ ($-\hat{\bmath{D}}$)
\citep[cf.][]{landau75}, i.e.,
$\hat{\bmath{u}}_+ \cdot \hat{\bmath{D}} > 0$ and $\hat{\bmath{u}}_- \cdot \hat{\bmath{D}} < 0$.
Then one has 
$\bmath{u}_+ \cdot \bmath{D} > 0$ and $\bmath{u}_- \cdot \bmath{D} < 0$ in the coordinate basis,
which lead to $\bmath{v}_+ \cdot \bmath{D} > 0$ and $\bmath{v}_- \cdot \bmath{D} < 0$ 
since $u^t > 0$. Therefore, the particle motions can carry $\bmath{J}_p \parallel \bmath{D}$.
This implies $\bmath{J}_p \perp \bmath{B}_p$. 
The force-free approximation $F_{\mu\nu} I^\nu = 0$ 
(i.e., $\rho \bmath{E} + \bmath{J} \times \bmath{B} = 0$) is violated in this case, since
$E_\varphi = 0$.

In summary, $D^2 > B^2$ is the necessary and sufficient condition for driving the electric
currents to flow across the poloidal $\bmath{B}$ field lines (and then obtaining $H_\varphi \neq 0$) 
under our assumptions (1)--(3) listed in Section~\ref{sec:BHmag1}.

\section{Unipolar Induction of Kerr Black Holes}
\label{sec:BHuni}

We discuss the unipolar induction process in the Kerr BH magnetosphere where the poloidal
$\bmath{B}$ field lines threading the ergosphere are open, i.e., crossing 
the outer light surface. We mainly utilize the BL coordinates rather than the KS ones
in this section.

The light surfaces are defined as follows.
In the BL coordinates, when $\hat{B}_{\varphi} = B_{\varphi} = 0$, 
the coordinate angular velocity of the drift motion is $\Omega_{\rm F}$, as deduced in Appendix~\ref{sec:app2}.
The light surfaces are the surfaces where
the four-velocity of a particle which is rotating with the coordinate angular velocity 
$\Omega_{\rm F}$ becomes null, i.e., $f(\Omega_{\rm F}, r, \theta) = 0$, where we have defined
\begin{equation}
f(\Omega_{\rm F}, r, \theta) \equiv (\xi + \Omega_{\rm F} \chi)^2 =
-\alpha^2 + \gamma_{\varphi \varphi} (\Omega_{\rm F} - \Omega)^2.
\label{eq:f_def}
\end{equation}
There can be two light surfaces. One has 
$\Omega_{\rm F} - \Omega = \alpha/\sqrt{\gamma_{\varphi\varphi}}$ at the outer light surface
(where $\Omega < \Omega_{\rm F}$),
while $\Omega_{\rm F} - \Omega = -\alpha/\sqrt{\gamma_{\varphi\varphi}}$ at the inner light surface
(where $\Omega > \Omega_{\rm F}$).

From the condition $\bmath{D} \cdot \bmath{B}=0$, 
equations~(\ref{eq:relation_E}) and (\ref{eq:omegaF_def}) lead to
\begin{equation}
\bmath{D} = -\frac{1}{\alpha} (\bmath{\omega} + \bmath{\beta}) \times \bmath{B}.
\label{eq:D}
\end{equation}
Calculating $D^2 \equiv D^i D_i$, one obtains in the BL coordinates
\begin{equation}
D^2 = \frac{1}{\alpha^2} (\Omega_{\rm F} - \Omega)^2 B_p^2,
\label{eq:basic0}
\end{equation}
where $B_p^2 \equiv B^r B_r + B^\theta B_\theta$. This equation can be rewritten by using
equation~(\ref{eq:f_def}) \citep[cf.][]{komissarov04}
\begin{equation}
(B^2 - D^2)\alpha^2 = -B^2 f(\Omega_F, r, \theta) + \frac{1}{\alpha^2} (\Omega_F - \Omega)^2 H_\varphi^2,
\label{eq:basic}
\end{equation}
where we have used $H_\varphi = \alpha B_\varphi$ (see equation~\ref{eq:relation_H}). 
This equation is useful for the following discussion.

\subsection{Origin of the electromotive force}
\label{sec:active}

We show that there is no steady, axisymmetric state with $\Omega_{\rm F} = 0$ and 
$H_\varphi = 0$ for the $\bmath{B}$ field lines threading the ergosphere. 
The condition $\Omega_{\rm F} = 0$ means $\bmath{E} = 0$, and
$\bmath{D} = (-1/\alpha) \bmath{\beta} \times \bmath{B}$. Equation~(\ref{eq:basic}) is reduced to 
\begin{equation}
(B^2 - D^2) \alpha^2 = B^2 (\alpha^2 - \beta^2).
\label{eq:OmegaF0case}
\end{equation}
In the ergosphere, where $\alpha^2 - \beta^2 < 0$, one has $D^2 > B^2$. Note that 
$B^2 - D^2 = F_{\mu\nu} F^{\mu\nu}/2$ is a scalar, and thus $B^2 - D^2 < 0$ is valid 
also in the KS coordinates.
The $\bmath{D}$ field stronger than the $\bmath{B}$ field drives the poloidal currents
to flow across the poloidal $\bmath{B}$ field lines, as discussed in Section~\ref{sec:particle}.
These poloidal currents generate $H_\varphi$ (equation~\ref{eq:AM2}).
Therefore, the state with $\Omega_{\rm F} = 0$ and $H_\varphi = 0$ cannot be maintained.

The poloidal currents (i.e., the charged particle flows) across the poloidal $\bmath{B}$ field lines 
due to the strong $\bmath{D}$ field in the ergosphere change the charge density distribution.
This reduces the strength of the $\bmath{D}$ field.
(We show the charge density distribution calculated for the Wald $\bmath{B}_p$ field
in Appendix~\ref{sec:app3},
which might be helpful for understanding the reduction of the $\bmath{D}$ field strength by
the charged particle flows across the poloidal $\bmath{B}$ field lines.)
Then equation~(\ref{eq:basic0}) implies that $\Omega_{\rm F} > 0$ is realized
and one finds a non-zero $\bmath{E}$ field.

From this argument, we can conclude that
the origin of the electromotive force is ascribed to the ergosphere in the unipolar induction
of the BH magnetosphere with $\bmath{D} \cdot \bmath{B} = 0$.

\begin{figure}
\begin{center}
\includegraphics[scale=0.6]{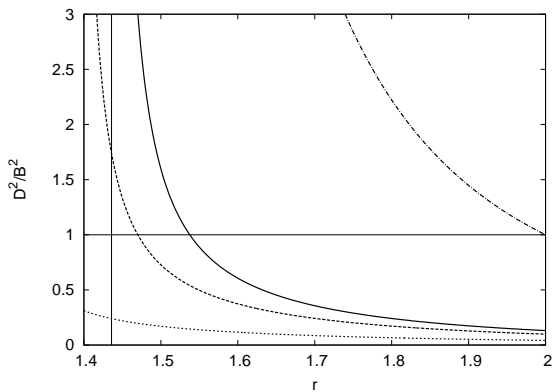}
\end{center}
\caption{
$D^2/B^2$ calculated for the Wald vacuum solution in the BL coordinates, as functions of 
$r$ for $\theta = \pi/2$ (solid), $0.45 \pi$ (dashed), and $0.4 \pi$ (dotted). The spin 
parameter is set as $a=0.9$. For comparison, $D^2/B^2$ ($=\beta^2/\alpha^2$) in the plasma-filled case with 
$\Omega_{\rm F} = 0$ and $H_\varphi = 0$ for $\theta = \pi/2$
is plotted by the dot-dashed line.
The vertical line represents the event horizon radius $r_{\rm H}=1.436$. 
The outer boundary of the ergosphere is $r_{\rm es} = 2$ for $\theta = \pi/2$.
}
\label{fig:dbbl}
\end{figure}

The generation of such a strong $\bmath{D}$ field may be understood as a phenomenon similar
to the pulsar case.
In the vacuum case, we can straightforwardly calculate $D^2/B^2$ by using the Wald solution
(see Appendix~\ref{sec:app1}), and find that the region where $D^2 > B^2$ is only
in the vicinity of the event horizon at the equatorial plane, as shown in Fig.~\ref{fig:dbbl}.
In contrast, one has $D^2/B^2 = \beta^2/\alpha^2$ in the plasma-filled case with $\Omega_{\rm F} = 0$ 
and $H_\varphi = 0$ (equation~\ref{eq:OmegaF0case}), which is larger than unity in the entire
ergosphere (see the dot-dashed line in Fig.~\ref{fig:dbbl}).
The enhancement of the electric field in the plasma-filled case compared to the vacuum case
is quite similar to the pulsar case (see equations~\ref{eq:pulsarE1} and \ref{eq:pulsarE2} 
in Section~\ref{sec:pulsar_potential}).
The charge distribution of the plasma screen the $\bmath{D}$ field component along the 
$\bmath{B}$ field but enhances the total strength of the $\bmath{D}$ field.
Note that the condition $D^2 > B^2$ is not due to a shortage of the number of charged particles
like the gap with non-zero electric field {\it along} the magnetic field lines 
\citep{blandford77,beskin92,hirotani98}, but rather, it arises due to a sufficiency of the 
charged particles sustaining $\bmath{D} \cdot \bmath{B} = 0$. 

The divergence of $D^2/B^2$ near the event horizon in Fig.~\ref{fig:dbbl} is not physical, 
just due to the BL coordinate singularity. In Appendix~\ref{sec:app1}, we calculate $D^2/B^2$ 
in the KS coordinates, which does not show any divergence (see Fig.~\ref{fig:db}). 

We also find that no $\bmath{B}$ field lines can have the condition of
$\Omega_{\rm F} = 0$ and $H_\varphi \neq 0$. Along such $\bmath{B}$ field lines, 
the poloidal electromagnetic angular momentum flux is non-zero, but the poloidal Poynting flux 
is zero (see equations~\ref{eq:AM2} and \ref{eq:energy2}). 
The current closure requires that such a $\bmath{B}$ field line should have a part where
the poloidal currents cross this field line. 
Focusing on the currents crossing the field line at the far zone, one finds that
the $\bmath{J}_p \times \bmath{B}_\varphi$ force
acts on the matter, converting the poloidal momentum flux of the electromagnetic field to that of 
the matter.
The matter should also gain the poloidal energy flux from the electromagnetic field, which
conflicts with the condition of $\Omega_{\rm F} = 0$. Therefore, the state with 
$\Omega_{\rm F} = 0$ and $H_\varphi \neq 0$ must not be realized.

\subsection{Maintenance of the poloidal currents}
\label{sec:maintenance}

As discussed above, the condition $\Omega_{\rm F} > 0$ is inevitably satisfied for all the 
$\bmath{B}$ field lines threading the ergosphere in the steady, axisymmetric state.
This condition leads to the azimuthal drift motions of the particles at the far zone, similar
to the pulsar case. For the {\it open} $\bmath{B}$ field lines that we consider, they must 
have the condition $B_{\varphi} = H_\varphi/\alpha \neq 0$ so that the azimuthal drift speeds 
do not exceed the light speed. Therefore, the open $\bmath{B}$ field lines threading the 
ergosphere will be forced to have the condition $\Omega_{\rm F} > 0$ and $H_\varphi \neq 0$.

We have seen that $H_\varphi$ is generated by the poloidal currents driven to flow across the poloidal
$\bmath{B}$ field line. Then the field line should have a part where $D^2 > B^2$ in the ergosphere (see also
Section~\ref{sec:structure} below) in order to maintain the 
poloidal currents, although those currents
reduce the strength of the $\bmath{D}$ field (leading to a larger value of $\Omega_{\rm F}$).
Therefore, the value of $\Omega_{\rm F}$ will be regulated so that $\Omega_{\rm F} > 0$ and the $\bmath{B}$
field line keeps having a part where $D^2 > B^2$.

It should be noted that the ergosphere is causally connected to the region around the outer light
surface, which is inside the outer fast magnetosonic surface \citep[cf.,][]{punsly08,beskin10,toma12}.
Thus the steady state with $\Omega_{\rm F} > 0$ and $H_\varphi \neq 0$ can be established in principle,
although more quantitative analyses on the stability of this state are needed.

If we have any $\bmath{B}$ field lines which penetrate the ergosphere but are not open,
e.g., field lines having the cylindrical shapes and not crossing the outer
light surfaces, they can be in the steady state with $\Omega_{\rm F} > 0$ and $H_\varphi = 0$.
Such field lines do not have a region where $D^2 > B^2$ (i.e., $\Omega_{\rm F}$ is larger
than the case of $H_\varphi \neq 0$).

\subsection{$\bmath{B}$ fields threading the equatorial plane}
\label{sec:structure}

\begin{figure}
\begin{center}
\includegraphics[scale=0.45]{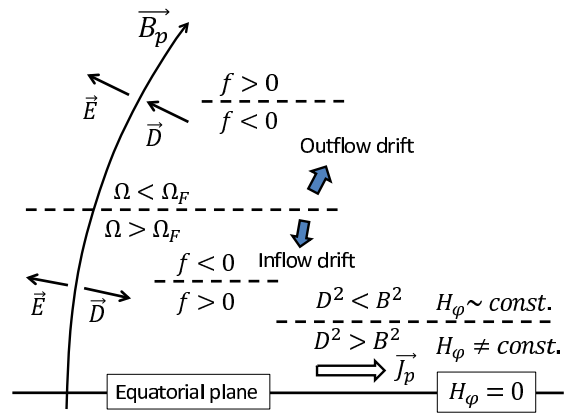}
\end{center}
\caption{
Electromagnetic structure in the northern hemisphere along a $\bmath{B}$ field line threading
the equatorial plane in the ergosphere.
The direction of the $\bmath{D}$ field changes at the point of $\Omega = \Omega_{\rm F}$.
This corresponds to the change of the $\bmath{D} \times \bmath{B}_{\varphi}$ drift direction.
The current crossing region is the region of $D^2 > B^2$, where $H_\varphi$ is not constant,
and located below the inner light surface ($f(\Omega_{\rm F}, r, \theta) = 0$).
}
\label{fig:structure}
\end{figure}

Let us call the region where $D^2 > B^2$ and then the poloidal currents are driven to
flow across the poloidal $\bmath{B}$ field line ``current crossing region."
In the case of $\Omega_{\rm F} =0$ and $H_\varphi = 0$, 
the current crossing region has been found to be the ergosphere in Section~\ref{sec:active}. 
Where is the current crossing region in the case of $\Omega_{\rm F} > 0$ and $H_\varphi \neq 0$?
Here we focus on an open $\bmath{B}$ field line {\it threading the equatorial plane}
in the ergosphere, showing a self-consistent electromagnetic structure along such a field line
and pinning down the region where $D^2 > B^2$.

First, since $\Omega_{\rm F} > 0$, there is a point where $\Omega = \Omega_{\rm F}$ for
each field line. 
One has $\bmath{D} \cdot \bmath{E} < 0$ below this point, while
$\bmath{D} \cdot \bmath{E} > 0$ above this point, as illustrated in Fig.~\ref{fig:structure}.
The point of $\Omega = \Omega_{\rm F}$ is located between the outer light surface
(where $\Omega_{\rm F} - \Omega = \alpha/\sqrt{\gamma_{\varphi\varphi}} > 0$)
and the inner light surface (where $\Omega_{\rm F} - \Omega = -\alpha/\sqrt{\gamma_{\varphi\varphi}} < 0$).
We plot the contour of $\Omega - \alpha/\sqrt{\gamma_{\varphi\varphi}}$ in Fig.~\ref{fig:fun},
which shows that $\Omega - \alpha/\sqrt{\gamma_{\varphi\varphi}}$ is positive only in the 
ergosphere. Thus $\Omega_{\rm F} > 0$ requires that the inner light surface is located in the ergosphere. 

Because of the symmetry, $H_\varphi = 0$ on the equatorial plane. The current crossing region
must include the equatorial plane, and $H_\varphi$ is generated in this region (see equation~\ref{eq:AM2}).
Equation~(\ref{eq:basic}) implies that $D^2 > B^2$ is satisfied in the region of 
$f(\Omega_{\rm F}, r, \theta) > 0$. Therefore, {\it the current crossing region must be below the 
inner light surface, which is within the ergosphere.} 
Above the current crossing region, $D^2 < B^2$ is satisfied and $H_\varphi \sim {\rm const}$.
(Here we do not consider the region where the currents flow across the poloidal $\bmath{B}$ field lines in the direction
of $\bmath{E}$, i.e., the particle acceleration zone.)
We summarize the electromagnetic structure in Fig.~\ref{fig:structure}.

In the current crossing region, one has $\bmath{D} \cdot \bmath{E} < 0$, so that
$(\bmath{J}_p \times \bmath{B}_p) \cdot \bmath{m} < 0$ and $\bmath{E} \cdot \bmath{J}_p < 0$,
which generate the poloidal electromagnetic angular momentum and energy fluxes, as seen
in equations~(\ref{eq:AM2}) and (\ref{eq:energy2}). 
Similar to the pulsar case (discussed in Section~\ref{sec:pulsar_current}), 
the poloidal currents are driven to flow across the poloidal $\bmath{B}$ field 
lines in the direction of $-\bmath{E}$ in the region where the outward energy flux is generated.

In the main body of the open field line region in the northern (southern) hemisphere, 
one has $H_\varphi < 0 (>0)$, i.e., the poloidal currents flow downward (upward).
The poloidal particle drift motions in the coordinate basis (which is in the direction of 
$\bmath{D} \times \bmath{B}_\varphi$ as shown in Appendix~\ref{sec:app2}) 
are directed towards infinity in the region of 
$\Omega < \Omega_{\rm F}$, while towards the equatorial plane in the region of 
$\Omega > \Omega_{\rm F}$.
The particles which drift towards the equatorial plane do not pass through the plane but
turn to move parallel to the plane due to the strong $\bmath{D}$ field.

\begin{figure}
\begin{center}
\includegraphics[scale=0.6]{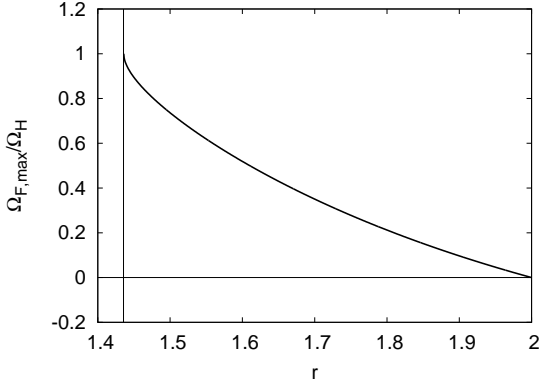}
\end{center}
\caption{
$\Omega_{\rm F,max}/\Omega_{\rm H}$ for the $\bmath{B}$ field lines threading
the equatorial plane in the ergosphere as functions of $r$ on the equatorial plane.
The spin parameter is set as $a=0.9$.
The vertical line represents $r_{\rm H}=1.436$. The radius of 
the outer boundary of the ergosphere is $r_{\rm es}=2$.
}
\label{fig:omega}
\end{figure}

\subsection{Range of $\Omega_{\rm F}$ and outflow power}
\label{sec:range}

A necessary condition for maintaining the electromagnetic structure along a $\bmath{B}$ field
line threading the equatorial plane in the ergosphere (Fig.~\ref{fig:structure})
is that the size of the current crossing region is finite. In the limit of the 
infinitesimally thin current crossing region, $D^2 \approx B^2$ on the equatorial plane.
Since $H_\varphi = 0$ there, equation~(\ref{eq:basic}) implies $f(\Omega_{\rm F}, r, \theta) \approx 0$.
This condition provides the maximum allowed value of $\Omega_{\rm F}$, i.e., we find the 
range of possible value of $\Omega_{\rm F}$ as 
\begin{equation}
0 < \Omega_{\rm F} < \Omega_{\rm F,max} \equiv \left. \left(\Omega - \frac{\alpha}{\sqrt{\gamma_{\varphi\varphi}}}
\right)\right|_{z=0}.
\label{eq:omegaF_cal}
\end{equation}
The value of $\Omega_{\rm F,max}$ normalized by the BH angular velocity $\Omega_{\rm H} = a/(2 r_{\rm H})$
as a function of $r$ on the equatorial plane is plotted in Fig.~\ref{fig:omega}. 
Since the poloidal current flows reduce the strength of the $\bmath{D}$ field, as discussed above,
$\Omega_{\rm F}$ is expected to have a value near $\Omega_{\rm F,max}$.

Fig.~\ref{fig:omega} implies that $\Omega_{\rm F}$ should be zero for the $\bmath{B}$ field line threading the ergosphere
at $r = r_{\rm es}$, which we call ``last ergospheric field line."
The poloidal currents will not cross this field line, and thus $H_\varphi = 0$
for this field line. The poloidal current circuit can 
consist of the current flow in the current crossing region, the outward flow along the 
last ergospheric field line, and the inward flow in the main body of the open field line region
(See Fig.~\ref{fig:current} and Section~\ref{sec:discussion}).

Similar to the pulsar case, it is reasonable that $|\bmath{B}_\varphi| = 
|B_\varphi|/\sqrt{\gamma_{\varphi\varphi}} \sim |\bmath{B}_p|$
around the outer light surface. Then we can have a rough estimate of the poloidal Poynting flux as 
$|\bmath{S}_p| = \Omega_{\rm F} |H_\varphi| |\bmath{B}_p|/4\pi 
\sim B_{p,{\rm ls}}^2/(4\pi)$, where we have used the equations
valid at the outer light surface $f(\Omega_{\rm F}, r, \theta) = -\alpha^2 + 
\gamma_{\varphi\varphi} (\Omega_F - \Omega)^2 = 0$, $\alpha \sim 1$, and $\Omega \ll \Omega_{\rm F}$.
When the open poloidal magnetic field roughly scales as 
$B_{p,{\rm ls}} \sim B_{p,{\rm H}} (r_{\rm ls}/r_{\rm H})^{-2}$ (i.e., monopole-like rather
than dipole), where $B_{p,{\rm H}}$ is
the field strength in the vicinity of the BH and $r_{\rm ls}$ is the radius of the outer light
surface, and then one has the luminosity per solid angle as
\begin{equation}
\frac{dL}{d\Omega} \sim r_{\rm ls}^2 |\bmath{S}_p| \sim \frac{B_{p,{\rm H}}^2 
\Omega_{\rm F}^2 r_{\rm H}^4}{4\pi} 
\sim \frac{a^2}{16\pi} \left(\frac{\Omega_{\rm F}}{\Omega_{\rm H}}\right)^2 
B_{p,{\rm H}}^2 r_{\rm H}^2.
\end{equation}
Here $\Omega_{\rm F}/\Omega_{\rm H}$ is expected to be $\sim \Omega_{\rm F,max}/\Omega_{\rm H}$ 
(Fig.~\ref{fig:omega}) for the $\bmath{B}$ field lines threading the equatorial plane in the ergosphere.

\section{Summary and Discussion}
\label{sec:discussion}

We consider the Blandford-Znajek process as the steady unipolar induction process  
in the Kerr BH magnetosphere in which a collisionless plasma is filled so that 
$\bmath{D} \cdot \bmath{B} = 0$ is sustained and the energy density is dominated by the electromagnetic field. 
The origin of 
the electromotive force in this process is the ergosphere, unlike in the
pulsar case, in which the origin of the electromotive force is the rotation of the stellar matter
(see Section~\ref{sec:pulsar}). 
All the open $\bmath{B}$ field lines threading the ergosphere inevitably keep having a part where
the $\bmath{D}$ field (perpendicular to the $\bmath{B}$ field) is stronger than the $\bmath{B}$ field,
i.e., $D^2 > B^2$, which drives the poloidal currents to flow across the poloidal 
$\bmath{B}$ field lines ($H_\varphi \neq 0$; see equation~\ref{eq:AM2}) and give rise to the electromotive force, i.e., $\Omega_{\rm F} > 0$. 
In the current crossing region, the currents flow in the direction of $-\bmath{E}$, which 
generate the poloidal Poynting flux (see equation~\ref{eq:energy2}), similar to the pulsar case.
The condition $\bmath{J}_p \times \bmath{B}_p \neq 0$ in the current crossing region
implies that the force-free approximation cannot be assumed (see Section~\ref{sec:particle}).
Note that we only assume $\bmath{D}\cdot\bmath{B} = 0$, not utilizing the force-free
condition $\rho\bmath{E}+\bmath{J}\times\bmath{B}=0$ in our arguments.

\subsection{$\bmath{B}$ fields threading the horizon}

\begin{figure}
\begin{center}
\includegraphics[scale=0.45]{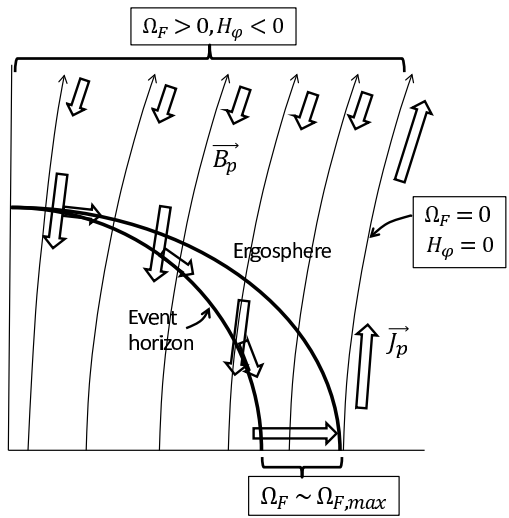}
\end{center}
\caption{
Expected electromagnetic structure of the BH magnetosphere (northern hemisphere). The thin solid arrows
represent the poloidal $\bmath{B}$ field lines, and the open arrows the poloidal currents. 
See the texts in Sections~\ref{sec:BHuni} and \ref{sec:discussion} for details.
}
\label{fig:current}
\end{figure}

We have shown the self-consistent electromagnetic structure along a $\bmath{B}$ field line threading
the equatorial plane in the ergosphere in Fig.~\ref{fig:structure}. Also for the $\bmath{B}$ field lines 
threading the event horizon, the above conclusion on the origin of the electromotive force
is applicable, and thus the electromagnetic 
structure will be similar to Fig.~\ref{fig:structure}, having the current crossing region just outside the horizon.
It is not an easy task 
to solve the whole electromagnetic structure of the BH magnetosphere, but our arguments so far 
allow us to conjecture that the solution looks like Fig.~\ref{fig:current}. 
We describe the poloidal current flows by the open arrows in Fig.~\ref{fig:current}.
Some fraction of the poloidal currents can flow across the horizon, while the remaining fraction
flows across the poloidal $\bmath{B}$ field lines just outside the horizon, generating the poloidal
Poynting flux.
In the current crossing region around the equatorial plane, the negatively charged particles 
flow across the horizon.
The current circuits will be made for the BH not to charge up in the steady state.

In the membrane paradigm \citep{thorne86,penna13}, the force-free condition
is assumed to be satisfied except on the membrane (at $r \approx r_{\rm H}$) 
covering the horizon, so that all the poloidal
currents flow into the membrane (i.e., flow across the horizon for freely falling observers)
in the region of the $\bmath{B}$ field lines threading the horizon.
The poloidal currents are then regarded as flowing along the viscous membrane and generating 
the outward Poynting flux. However, this picture looks unphysical, because the membrane is causally 
disconnected with its exterior \citep{punsly89,punsly08}. The energy equation (\ref{eq:energy2}) implies
that a region of $\bmath{E} \cdot \bmath{J}_p < 0$ has to be causally connected for producing the 
Poynting flux. Such a region is realized outside the horizon (or the membrane) and within the 
ergosphere, where the force-free condition is violated.

We have deduced the maximum value of $\Omega_{\rm F}$, $\Omega_{\rm F,max}$, for the poloidal $\bmath{B}$ field lines 
threading the equatorial plane in the ergosphere as shown in Fig.~\ref{fig:omega}. This 
can be deduced easily by utilizing the symmetry condition on the equatorial plane $H_\varphi = 0$
for equation (\ref{eq:basic}).
In contrast, for the poloidal $\bmath{B}$ field lines threading the horizon, 
$\Omega_{\rm F,max}$ is difficult to deduce, because $H_\varphi \neq 0$ at the horizon in general.
Furthermore, the BL coordinates are not useful for analyzing the condition near the horizon.
Equation (\ref{eq:D}) can be rewritten {\it in the KS coordinates} as
\begin{equation}
(B^2 - D^2)\alpha^2 = - B^2 f(\Omega_{\rm F}, r, \theta) + (\Omega_{\rm F}B_\varphi + \beta^r B_r)^2,
\end{equation}
where $B_\varphi = (H_\varphi + \sqrt{\gamma} \beta^r D^\theta)/\alpha$.
This implies that $\Omega_{\rm F,max}$, which is obtained for $D^2 \approx B^2$ at the horizon, depends on
the structure of $\bmath{B}_p$.

We have assumed that some field lines of $\bmath{B}_p$ produced by the external currents
are threading the equatorial plane within
the ergosphere in Sections~\ref{sec:structure} and \ref{sec:range} and in Fig.~\ref{fig:current}.
However, a situation could be realized in which
the magnetospheric currents around the equatorial plane become so strong that
no field lines intersect the equatorial plane within the ergosphere, as demonstrated
by some MHD numerical simulations \citep{komissarov05,komissarov07}. The Blandford-Znajek process can operate 
only for the field lines threading the horizon in this case. 

\subsection{On the ideal MHD condition}

The ideal MHD condition ($F_{\mu\nu} u^{\nu} = 0$) is $\hat{\bmath{D}} +
\hat{\bmath{v}}\times\hat{\bmath{B}}=0$ as viewed by FIDOs.
\citet{beskin_rafikov00} examined the two-fluid effects on the electron-positron plasma
in the pulsar magnetosphere with a monopole magnetic field and the number density much larger
than the Goldreich-Julian density, and showed that the one-fluid flow with the ideal MHD condition
is a good approximation in the region $E^2 < B^2$ \citep[see also][]{melatos96,goodwin04}.
In this case, the ideal MHD condition does not imply the infinite isotropic conductivity, but rather,
the flow is almost force-free and moving via the $\bmath{E}\times\bmath{B}$ drifts, i.e.,
$\bmath{J}_p \parallel \bmath{B}_p$. Most region in the BH magnetosphere may be described
as the one-fluid with the ideal MHD condition or the force-free plasma in the same sense.
However, this has to be violated in the current crossing region,
as stated above. (The ideal MHD condition requires $D^2<B^2$ from $|\hat{\bmath{v}}|<1$.)

For the pulsar case, the fluid at and inside the stellar surface has a high isotropic 
conductivity owing to the effective particle collisions, i.e.,
the particle collision rate is very high compared with the gyration period 
\citep[cf.][]{bekenstein78,khanna98}, in which the poloidal currents flow across the 
poloidal magnetic field lines and the electromotive force arises.
The ideal MHD condition is a good approximation there.

\citet{beskin00} examined the one-fluid MHD flow in the Kerr BH magnetosphere with a small
spin parameter $a \ll 1$ and a monopole magnetic field, and argued that there are many
types of solutions corresponding to different values of $H_\varphi$ and $\Omega_{\rm F}$ 
(see their Table 1). 
The MHD flow in the Kerr BH magnetosphere generally consists of an inflow and an outflow,
and they assumed that there is a gap between them, in which $\bmath{D}\cdot\bmath{B} \neq 0$
and the poloidal currents flow across the poloidal $\bmath{B}$ field lines. 
Some types of those solutions have a surface of $D^2 = B^2$ in the inflow, 
``inflow with shock" \citep[see also][]{beskin83,beskin_rafikov00}, 
which appear to be consistent with our conclusion, while
one of those solutions has no surfaces of $D^2 = B^2$ for which
the ideal MHD condition is satisfied in the whole region except the gap region.
Whether the latter type of the solution is realized depends on the nature of the gap and
possibly on the assumptions on $a$ and the magnetic field structure, and
this point remains to be further investigated.

\subsection{Final remarks}

The plasma may keep being injected into the whole Kerr BH magnetosphere through the electron-positron
pair creation by collisions of two photons emitted from the external hot fluid and/or through 
electron-proton creation by decays of neutrons escaping from the external fluid \citep{toma12}.
The electrons accelerated due to the strong $\bmath{D}$ field in the current crossing region
produce photons, which also can create electron-positron pairs by collisions with the external photons.
However, it will be hard for those electron-positron pairs to be distributed widely in the magnetosphere,
because the poloidal drift motions ($\bmath{D} \times \bmath{B}_\varphi$) are downward in the 
region of $\Omega > \Omega_{\rm F}$ (see Fig.~\ref{fig:structure}).

The origin of the electromotive force in the Blandford-Znajek process is attributed to
the ergosphere. It cannot work in the Schwarzschild BH magnetosphere, which has no
ergosphere. Then it is expected that the rotational energy of the Kerr BH decreases by this 
process. However, it is not evident how the BH rotational energy is converted to the Poynting
flux. This energy conversion mechanism has been discussed so far by utilizing the ideal MHD simulation 
results, focusing on the importance of generation of negative particle energies 
\citep{koide02,komissarov05}.
We have argued that the breakdown of the ideal MHD condition is
essential for giving rise to the electromotive force. Further analyses on 
the energy transport are needed to fully understand the physics of the 
Blandford-Znajek process
\citep[see also][]{komissarov09}.

\section*{Acknowledgments}

We thank the anonymous referee for his/her comments, which improved the manuscript.
K.T. thanks Shinpei Shibata,
Hideyuki Tagoshi, Yuto Teraki, and Shigeo Kimura for useful discussions. 
This work is partly supported by JSPS Research Fellowships for Young Scientists No.231446.

\appendix

\section{Electromagnetic Field in the Vacuum}
\label{sec:app1}

\cite{wald74} showed a vacuum test electromagnetic field solution in Kerr space-time for which
the magnetic field is uniform, parallel to the rotation axis, at infinity with the strength $B_0$ as
\begin{equation}
F_{\mu\nu} = B_0 (\partial_{[\mu} \xi_{\nu]} + 2a \partial_{[\mu} \chi_{\nu]}),
\end{equation}
where $a$ is the dimensionless spin parameter.
The covariant forms of the $\bmath{D}$ and $\bmath{B}$ fields are given by
\begin{equation}
D_\mu = F_{\mu \nu} n^\nu, ~~~ B^\mu = -{}^*F^{\mu \nu} n_\nu,
\label{eq:BDdef}
\end{equation}
\citep{komissarov04}.
In this sense, $\bmath{D}$ and $\bmath{B}$ are the electric and magnetic fields as measured by
FIDOs. Here we calculate the strengths of the $\bmath{D}$ and $\bmath{B}$ fields of the vacuum solution
in the BL and KS coordinates.

In the BL coordinates, one has the following non-zero
metric components
\begin{eqnarray}
&& \alpha = \sqrt{\frac{\varrho^2 \Delta}{\Sigma}}, ~~~ \beta^{\varphi} = -\frac{2ar}{\Sigma}, \nonumber \\
&& \gamma_{\varphi \varphi} = \frac{\Sigma}{\varrho^2} \sin^2 \theta, ~~~ \gamma_{rr} = \frac{\varrho^2}{\Delta}, ~~~
\gamma_{\theta \theta} = \varrho^2, 
\end{eqnarray}
where
\begin{eqnarray}
&& \varrho^2 = r^2 + a^2 \cos^2 \theta, ~~~ \Delta = r^2 + a^2 - 2r, \nonumber \\
&& \Sigma = (r^2 + a^2)^2 - a^2 \Delta \sin^2 \theta,
\end{eqnarray}
\citep[see e.g.][]{landau75}.
FIDOs can have a local orthonormal basis as
\begin{eqnarray}
&& \omega^t = \alpha dt, ~~~ \omega^\varphi = \sqrt{\gamma_{\varphi \varphi}} (\beta^\varphi dt + d\varphi), 
\nonumber \\
&& \omega^r = \sqrt{\gamma_{rr}}dr, ~~~ \omega^\theta = \sqrt{\gamma_{\theta \theta}} d\theta,
\end{eqnarray}
which satisfy $ds^2 = \eta_{ab} \omega^a \omega^b$. The corresponding basis vectors are
\begin{eqnarray}
&& e_t = \frac{1}{\alpha} (\partial_t - \beta^\varphi \partial_\varphi), ~~~
e_\varphi = \frac{1}{\sqrt{\gamma_{\varphi \varphi}}} \partial_\varphi, \nonumber \\
&& e_r = \frac{1}{\sqrt{\gamma_{rr}}} \partial_r, ~~~ e_\theta = \frac{1}{\sqrt{\gamma_{\theta \theta}}} \partial_\theta.
\end{eqnarray}
Note that $n^\mu = e_t^\mu$ and $n_\mu = -\omega^t_\mu$. Then we can calculate the vector components
$\hat{D}_i$ and $\hat{B}^i$ in respect of the orthonormal basis as
\begin{equation}
\hat{D}_i = D_\mu e^\mu_i = F_{\mu \nu} e^\mu_i e^\nu_t, ~~~
\hat{B}^i = B^\mu \omega_\mu^i = \frac{1}{2} e^{\mu \nu \alpha \beta} F_{\alpha \beta} \omega_\mu^i \omega_\nu^t.
\label{eq:FIDO_DB}
\end{equation}
The result is
\begin{eqnarray}
&&\hat{D}_{\varphi} = \hat{B}^\varphi = 0, \nonumber \\
&&\hat{D}_r = \nonumber \\
&&\frac{B_0 a}{\sqrt{\Sigma} \rho^4} \left[2r^2 \rho^2 \sin^2\theta -
(r^2+a^2)(r^2-a^2 \cos^2\theta) (1+\cos^2\theta)\right], \nonumber \\
&&\hat{D}_\theta = \frac{B_0 a^3 \sqrt{\Delta}}{\sqrt{\Sigma} \rho^4}
2 r \sin\theta \cos\theta (1+\cos^2\theta), \label{eq:BLfield} \\
&&\hat{B}^r = \frac{B_0}{\sqrt{\Sigma} \rho^4} \cos\theta \left[(r^2+a^2)
(\rho^4-4ra^2\cos^2\theta) - 2ra^4 \sin^4\theta \right], \nonumber \\
&&\hat{B}^\theta = \frac{-B_0 \sqrt{\Delta}}{\sqrt{\Sigma} \rho^4} \sin\theta \left[
r\rho^4 + a^2 (r^2-a^2\cos^2\theta)(1+\cos^2\theta)\right]. \nonumber
\end{eqnarray}
This agrees with the result of the same calculation in \cite{king75} and \cite{punsly89}. 
Because $D^\mu D_\mu$ and $B^\mu B_\mu$ are scalars and $\hat{D}^t = D^t = 0$
and $\hat{B}^t = B^t = 0$, one has $D^2 = \hat{D}^2$ and $B^2 = \hat{B}^2$.
We calculate $\hat{D}^2/\hat{B}^2 = D^2/B^2$ and plot it in Figure~\ref{fig:dbbl}.

In the KS coordinates, one has the following non-zero metric components
\begin{eqnarray}
&&\alpha = \frac{1}{\sqrt{1+z}}, ~~~ \beta^r = \frac{z}{1+z}, ~~~ 
\gamma_{\varphi\varphi}=\frac{\Sigma}{\rho^2}\sin^2\theta, \nonumber \\
&&\gamma_{r\varphi} = -a (1+z)\sin^2\theta, ~~~ \gamma_{rr} = 1+z, 
~~~\gamma_{\theta\theta}=\varrho^2
\end{eqnarray} 
where $z = 2r/\varrho^2$ \citep{komissarov04}. By transferring the KS FIDO motion into the BL 
coordinates, one finds that the KS FIDO has the same angular velocity as the BL FIDO but also
moves radially towards the true singularity. The orthonormal basis dual-vectors can be chosen as
\begin{eqnarray}
&& \omega^t = \alpha dt, ~~~
\omega^\varphi = \beta^r \frac{\gamma_{r\varphi}}{\sqrt{\gamma_{\varphi \varphi}}}dt
+\sqrt{\gamma_{\varphi \varphi}}d\varphi + \frac{\gamma_{r\varphi}}{\sqrt{\gamma_{\varphi\varphi}}}dr,
\nonumber \\
&& \omega^r = \sqrt{\frac{\gamma}{\gamma_{\varphi \varphi} \gamma_{\theta\theta}}} (\beta^r dt +dr), ~~~
\omega^\theta = \sqrt{\gamma_{\theta\theta}} d\theta,
\end{eqnarray}
and the corresponding basis vectors are
\begin{eqnarray}
&& e_t = \frac{1}{\alpha} \left(\partial_t - \beta^r \partial_r \right), ~~
e_\varphi = \frac{1}{\sqrt{\gamma_{\varphi\varphi}}} \partial_\varphi, ~~ \nonumber \\
&& e_r = \sqrt{\frac{\gamma_{\varphi\varphi}\gamma_{\theta\theta}}{\gamma}}
\left(\partial_r - \frac{\gamma_{r\varphi}}{\gamma_{\varphi\varphi}}\partial_\varphi \right), ~~
e_\theta = \frac{1}{\sqrt{\gamma_{\theta\theta}}}\partial_{\theta},
\end{eqnarray}
\citep{komissarov04b}. Note that $n^\mu = e_t^\mu$ and $n_\mu = -\omega^t_{\mu}$, similar to those in
the BL coordinates. Then by calculating equation~(\ref{eq:FIDO_DB}) we obtain
\begin{eqnarray}
&& \hat{D}_{\varphi} = \nonumber \\
&& \frac{B_0}{\sqrt{\Sigma} \rho^5 \sqrt{1+z}} 2r \sin\theta \left[
r\rho^4 + a^2 (r^2-a^2\cos^2\theta)(1+\cos^2\theta)\right], \nonumber \\
&&\hat{D}_r = \nonumber \\
&&\frac{B_0 a}{\sqrt{\Sigma}\rho^4} \left[2r^2 \rho^2 \sin^2\theta -
(r^2+a^2)(r^2-a^2 \cos^2\theta) (1+\cos^2\theta)\right], \nonumber \\
&&\hat{D}_{\theta} = \frac{B_0 a^3}{\rho^5 \sqrt{1+z}} 2r \sin\theta \cos\theta (1+\cos^2\theta), \label{eq:KSfield}\\
&&\hat{B}^{\varphi} = \frac{B_0 a^3}{\sqrt{\Sigma} \rho^5 \sqrt{1+z}} 4r^2
\sin\theta \cos\theta (1+\cos^2\theta), \nonumber \\
&&\hat{B}^r = \frac{B_0}{\sqrt{\Sigma}\rho^4} \cos\theta \left[(r^2+a^2)
(\rho^4-4ra^2\cos^2\theta) - 2ra^4 \sin^4\theta \right], \nonumber \\
&&\hat{B}^\theta = \frac{-B_0}{\rho^5\sqrt{1+z}} \sin\theta \left[
r\rho^4 + a^2 (r^2-a^2\cos^2\theta)(1+\cos^2\theta)\right]. \nonumber
\end{eqnarray}

\begin{figure}
\begin{center}
\includegraphics[scale=0.6]{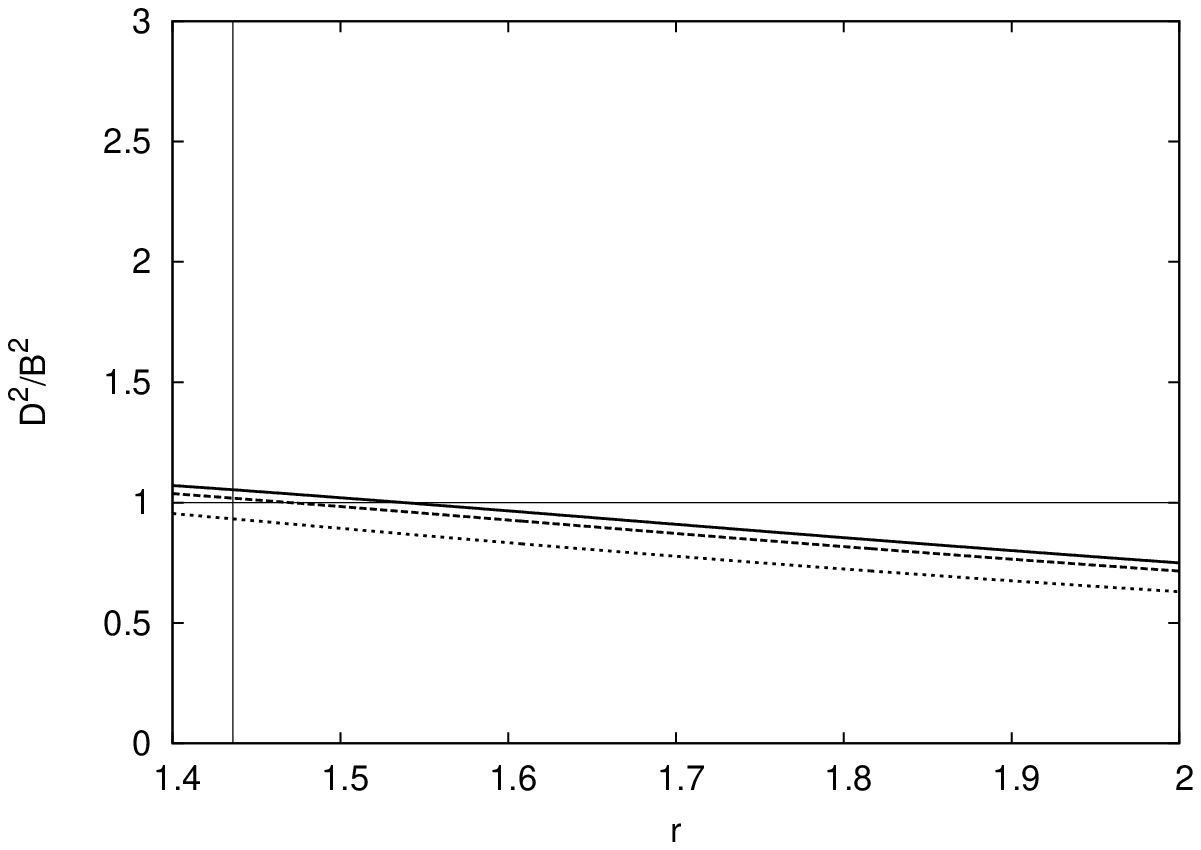}
\end{center}
\caption{
$D^2/B^2$ calculated for the Wald vacuum solution in the KS coordinates, as functions of 
$r$ for $\theta = \pi/2$ (solid), $0.45 \pi$ (dashed), and $0.4 \pi$ (dotted). The spin 
parameter is set as $a=0.9$. 
The vertical line represents the event horizon radius $r_{\rm H}=1.436$. 
The radius of the outer boundary of the ergosphere is $r_{\rm es} = 2$ for $\theta = \pi/2$.
}
\label{fig:db}
\end{figure}

In Fig.~\ref{fig:db} we plot $D^2/B^2 = \hat{D}^2/\hat{B}^2$ calculated in the KS coordinates.
It does not show the divergence at the event horizon, unlike that calculated in the BL coordinates
as shown in Fig.~\ref{fig:dbbl}. Note that the $\bmath{D}$ fields (as well as the $\bmath{B}$ fields)
in the BL and KS coordinates are not identical, correspondingly to the difference of $n^{\mu}$ in
equation~(\ref{eq:BDdef}). Thus $D^2$, $B^2$, $D^2/B^2$ are different in the two coordinates, but 
$B^2 - D^2 = F_{\mu\nu} F^{\mu\nu}/2$ is a scalar, providing the same value in the two
coordinates. This can be confirmed directly by using equations~(\ref{eq:BLfield}) and (\ref{eq:KSfield}).
As a result, the region of $(r, \theta)$ where $D^2/B^2 > 1$ (i.e., $D^2 - B^2 > 0$) are identical
in Fig.~\ref{fig:dbbl} and Fig.~\ref{fig:db}.

\section{Particle Motion as Viewed by FIDOs}
\label{sec:app2}

The equation of a particle motion as viewed by FIDOs is described as
\begin{equation}
\frac{D \hat{u}_\mu}{d\tau} = \frac{q}{m} \hat{F}_{\mu \nu} \hat{u}^\nu,
\end{equation}
where $\tau$, $q$, and $m$ are the proper time, charge, and mass of a particle, respectively. 
For $\mu = \varphi$ in the BL coordinates 
as an example, one has
\begin{eqnarray}
\hat{F}_{\varphi t} &=& F_{\alpha \beta} e_\varphi^\alpha e_t^\beta = \hat{D}_\varphi, \nonumber \\
\hat{F}_{\varphi r} &=& F_{\alpha \beta} e_\varphi^\alpha e_r^\beta \nonumber \\
&=& e_{\varphi r \theta} B^\theta
\frac{1}{\sqrt{\gamma_{\varphi\varphi}}} \frac{1}{\sqrt{\gamma_{rr}}} = \sqrt{\gamma_{\theta\theta}} 
B^\theta = \hat{B}^\theta, \nonumber \\
\hat{F}_{\varphi \theta} &=& F_{\alpha \beta} e_\varphi^\alpha e_\theta^\beta \nonumber \\
&=& e_{\varphi \theta r} B^r
\frac{1}{\sqrt{\gamma_{\varphi\varphi}}} \frac{1}{\sqrt{\gamma_{\theta\theta}}} = -\sqrt{\gamma_{rr}} 
B^r = -\hat{B}^r, \nonumber
\end{eqnarray}
where we have used $F_{ij} = e_{ijk} B^k$ \citep[see][]{komissarov04}. 
Thus one obtains
\begin{equation}
\frac{D\hat{u}_{\varphi}}{d\tau} = \frac{q}{m} (\hat{D}_{\varphi} \hat{u}^t + \epsilon_{\varphi jk} 
\hat{u}^j \hat{B}^k).
\end{equation}
The same form of equation is obtained also for $\mu = r$ and $\theta$. By taking account of 
$d\tau = d\hat{t}/\hat{u}^t$ and the assumption that the gravitational force is 
negligible compared with the Lorentz force, one obtains equation (\ref{eq:EOM}).
The same form of the equation
is obtained also by using the KS coordinates.

Let us examine the drift motion of a particle in the plasma with $\bmath{D} \cdot \bmath{B} = 0$
in the BL coordinates.
From equation~(\ref{eq:D}), one has $D_r = (\sqrt{\gamma}/\alpha)(\Omega_{\rm F} - \Omega)B^\theta$,
$D_\theta = (-\sqrt{\gamma}/\alpha)(\Omega_{\rm F} - \Omega)B^\theta$, and $D_\varphi = 0$.
The drift three-velocity as viewed by FIDOs is $\hat{v}_d = (\hat{\bmath{D}} \times
\hat{\bmath{B}})/\hat{B}^2$, where $\hat{B}^2 = B^2$. 
The azimuthal drift velocity is calculated as
\begin{equation}
\hat{v}_d^\varphi = \frac{D_r B_\theta - D_\theta B_r}{B^2 \sqrt{\gamma_{rr}\gamma_{\theta\theta}}}
= \frac{\sqrt{\gamma_{\varphi\varphi}}}{\alpha}(\Omega_{\rm F} - \Omega) \frac{B_p^2}{B^2}.
\end{equation}
In general, one has $\hat{u}^t = u^\mu \omega_\mu^t = \alpha u^t$ and
$\hat{u}^\varphi = u^\mu \omega_\mu^\varphi = \sqrt{\gamma_{\varphi\varphi}}(u^\varphi-\Omega u^t)$.
Then $\hat{v}^\varphi = \hat{u}^\varphi/\hat{u}^t = 
(\sqrt{\gamma_{\varphi\varphi}}/\alpha)(v^\varphi - \Omega)$. Therefore, one finds
\begin{equation}
v_d^\varphi = \Omega_{\rm F}\left(1-\frac{B_\varphi B^\varphi}{B^2}\right) 
+ \Omega \frac{B_\varphi B^\varphi}{B^2}.
\end{equation}
When $\hat{B}_\varphi = B_\varphi = 0$, one has $v_d^\varphi = \Omega_{\rm F}$. In this case,
$\hat{v}_d^\varphi = \pm 1$ at the light surfaces, where $\Omega_{\rm F} - \Omega = \pm
\alpha/\sqrt{\gamma_{\varphi\varphi}}$.

When $B_\varphi \neq 0$, one has non-zero poloidal components of the drift velocity,
\begin{equation}
\hat{v}_d^r = \frac{D_\theta B_\varphi}{B^2 \sqrt{\gamma_{\theta\theta}\gamma_{\varphi\varphi}}}, ~~~
\hat{v}_d^\theta = \frac{-D_r B_\varphi}{B^2 \sqrt{\gamma_{rr}\gamma_{\varphi\varphi}}}.
\end{equation}
Then one has $v_d^r = \hat{v}_d^r/\sqrt{\gamma_{rr}} = (\bmath{D} \times \bmath{B}_\varphi)^r/B^2$
and $v_d^\theta = \hat{v}_d^\theta/\sqrt{\gamma_{\theta\theta}} 
= (\bmath{D} \times \bmath{B}_\varphi)^\theta/B^2$. This means that 
\begin{equation}
\bmath{v}_d^p \parallel (\bmath{D} \times \bmath{B}_\varphi).
\end{equation}

\section{Charge density distribution}
\label{sec:app3}

We derive a description of the charge distribution which is valid for 
$f(\Omega_{\rm F}, r, \theta) < 0$ in the case of $\bmath{D} \cdot \bmath{B}=0$ and $B_\varphi = 0$
in the BL coordinates.
By using equations~(\ref{eq:D}) and (\ref{eq:relation_H}), one has
\begin{eqnarray}
4\pi \rho &=& \nabla \cdot \bmath{D} = \nabla \cdot 
\left[\frac{-1}{\alpha} (\bmath{\omega} + \bmath{\beta}) \times \bmath{B}\right] \nonumber \\
&=& \frac{\gamma_{\varphi\varphi}(\Omega_{\rm F}-\Omega)}{\alpha^2 \sqrt{\gamma}}
(\partial_r H_\theta - \partial_\theta H_r) \nonumber \\
&& + \frac{1}{\sqrt{\gamma}} (H_\theta \partial_r - H_r \partial_\theta) 
\frac{\gamma_{\varphi\varphi}(\Omega_{\rm F} - \Omega)}{\alpha^2} \nonumber \\
&& -\frac{\gamma_{\varphi\varphi}(\Omega_{\rm F}-\Omega)\Omega}{\alpha^2} \nabla \cdot \bmath{D}
\nonumber \\
&& - \bmath{D} \cdot \nabla \left[\frac{\gamma_{\varphi\varphi}(\Omega_{\rm F}-\Omega)\Omega}
{\alpha^2}\right].
\end{eqnarray}
The particle drift motions carry the current as measured by FIDOs, 
$\hat{j}^\varphi = \rho \hat{v}^\varphi_d$,
which is equivalent to $j^\varphi = (\rho/\alpha)(v^\varphi_d - \Omega)$.
By using $J^\varphi = \alpha j^\varphi + \rho \Omega$ (Equation~\ref{eq:Jj}), one has
\begin{eqnarray}
\partial_r H_\theta - \partial_\theta H_r = 4\pi \sqrt{\gamma} J^\varphi
= 4\pi \sqrt{\gamma} \rho \Omega_{\rm F}.
\end{eqnarray}
Then one obtains (by using equation~\ref{eq:relation_H} again)
\begin{eqnarray}
\rho &=& \frac{\alpha^2}{4\pi f(\Omega_{\rm F}, r, \theta)}
\left[\frac{\gamma_{\varphi\varphi}(\Omega_{\rm F}-\Omega)}{\alpha^2}
(D^r \partial_r + D^\theta \partial_\theta)\Omega \right. \nonumber \\
&& \left. - \frac{\alpha}{\sqrt{\gamma}} (B_\theta \partial_r - B_r \partial_\theta)
\frac{\gamma_{\varphi\varphi}(\Omega_{\rm F}-\Omega)}{\alpha^2}\right]
\end{eqnarray}
This equation can be reduced to
\begin{eqnarray}
\rho &=& \frac{\alpha^2}{4\pi f(\Omega_{\rm F}, r, \theta)}
\left\{\frac{\gamma_{\varphi\varphi}}{\alpha\sqrt{\gamma}}
\left[\frac{\gamma_{\varphi\varphi}(\Omega_{\rm F}-\Omega)^2}{\alpha^2}+1\right]
(B_\theta \partial_r - B_r \partial_\theta)\Omega \right. \nonumber \\
&& \left. - \frac{\alpha}{\sqrt{\gamma}} (\Omega_{\rm F}-\Omega) 
(B_\theta \partial_r - B_r \partial_\theta)\frac{\gamma_{\varphi\varphi}}{\alpha^2}
- \frac{\gamma_{\varphi\varphi}}{\alpha} B_p^2 \frac{d\Omega_{\rm F}}{d\Psi}
\right\}
\end{eqnarray}
The charge density $\rho$ can be calculated if $\bmath{B}_p$ and $\Omega_{\rm F}(\Psi)$
are given.

\begin{figure}
\begin{center}
\includegraphics[scale=0.7]{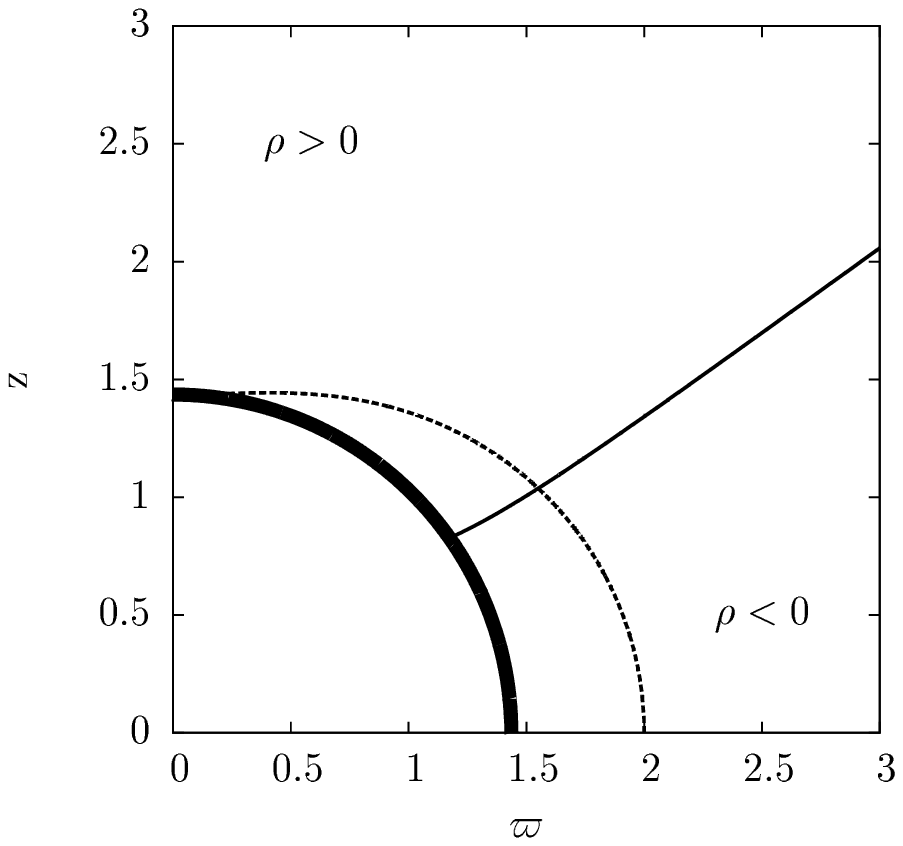}
\end{center}
\caption{
Charge density distribution for the Wald $\bmath{B}_p$ field with $\bmath{D} \cdot \bmath{B} = 0$ and
$\Omega_{\rm F} = 0$. The surface of $\rho = 0$ is represented by the 
thin black line. One has $\rho <0$ under this line and outside the event horizon
(the thick black line). The inner light surface is identical to the outer boundary 
of the ergosphere (the dashed line).
}
\label{fig:cdd11}
\end{figure}

Let us calculate $\rho$ by assuming $\bmath{B}_p$ as the Wald vacuum solution
(equation~\ref{eq:BLfield}). This corresponds to the Goldreich-Julian charge density for the 
Kerr BH magnetosphere.
For $\Omega_{\rm F} = \Omega_{\rm H}/2$, we confirmed that our calculation result of the 
surface of $\rho = 0$ is consistent with Fig. 3 of \citet{beskin92}.
For $\Omega_{\rm F} = 0$, we obtain the result as shown in Figure~\ref{fig:cdd11}.

\begin{figure}
\begin{center}
\includegraphics[scale=0.7]{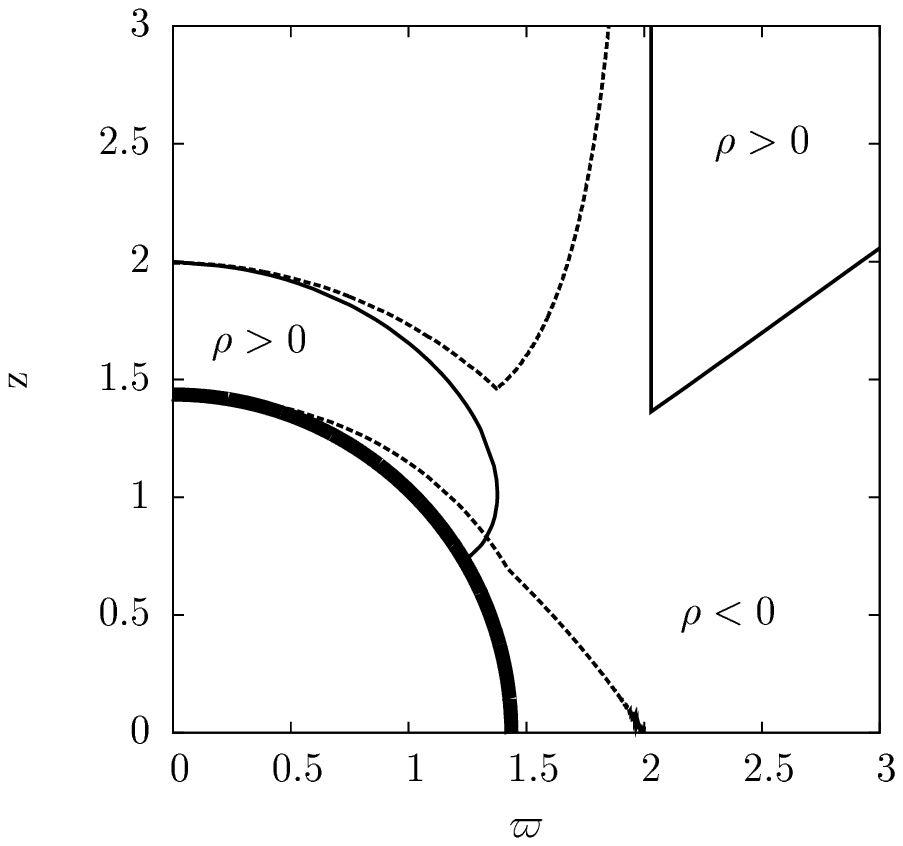}
\end{center}
\caption{
Charge density distribution for the Wald $\bmath{B}_p$ field with $\bmath{D} \cdot \bmath{B} = 0$ and
$\Omega_{\rm F}$ given by equation~(\ref{eq:omegaF_ref}). The surface of $\rho=0$ is represented
by the solid lines. In the region surrounded by the two solid lines and the event horizon
(the thick solid line), one has $\rho < 0$. The inner dashed line represents the inner light
surface, and the outer dashed line the surface of $\Omega = \Omega_{\rm F}$.
}
\label{fig:cddm2}
\end{figure}

For the case of non-uniform $\Omega_{\rm F}$, the flux function $\Psi$ is required to calculate $\rho$.
For the Wald solution of the $\bmath{B}$ field, it is given as \citep{beskin92}
\begin{equation}
\Psi = \frac{B_0 \sin^2\theta}{2\varrho^2} (\Sigma - 4a^2 r).
\end{equation}
It is confirmed that this form of $\Psi$ provides the $\bmath{B}$ field in equation~(\ref{eq:BLfield})
through equation~(\ref{eq:Psi}).

As an example, we assume the Wald solution of the $\bmath{B}_p$ field and $\Omega_{\rm F}$ given as
\begin{equation}
\Omega_{\rm F}(\Psi) = 
\left\{
\begin{array}{l}
\Omega_{\rm H}/2 ~~~~~~~ {\rm for}~\varpi_{\rm eq}(\Psi) \leq r_{\rm H} \\
\frac{\Omega_{\rm H}}{2} \frac{r_{\rm es} - \varpi_{\rm eq}(\Psi)}{r_{\rm es} - r_{\rm H}}
~~~ {\rm for}~r_{\rm H} < \varpi_{\rm eq}(\Psi) < r_{\rm es} \\
0 ~~~~~~~ {\rm for}~\varpi_{\rm eq}(\Psi) \geq r_{\rm es},
\end{array}
\right.
\label{eq:omegaF_ref}
\end{equation}
where $\varpi_{\rm eq}(\Psi)$ is the cylindrical radius at which a $\bmath{B}$ field line
crosses the equatorial plane. This form satisfies $\Omega_{\rm F} \la \Omega_{\rm F,max}$
for the $\bmath{B}$ field lines threading the equatorial plane outside the event horizon,
where $\Omega_{\rm F,max}$ is given by equation~(\ref{eq:omegaF_cal}) (see also Fig.~\ref{fig:omega}).
The calculation result is shown in Fig.~\ref{fig:cddm2}.

\bsp

\label{lastpage}


\begin{thebibliography}{99}
\bibitem[\protect\citeauthoryear{Asano \& Takahara}{2009}]{asano09} 
Asano K., \& Takahara F., 2009, ApJ, 690, L81
%
\bibitem[\protect\citeauthoryear{Bardeen et al.}{1972}]{bardeen72} 
Bardeen J.M., Press W.H., \& Teukolsky S.A., 1972, ApJ, 178, 347
%
\bibitem[\protect\citeauthoryear{Becker et al.}{2011}]{becker11} 
Becker P.A., Das S., \& Le T., 2011, ApJ, 743, 47
%
\bibitem[\protect\citeauthoryear{Bejger et al.}{2012}]{bejger12} 
Bejger M., Piran, T., Abramowicz M., \& Hakanson F. 2012, Phys. Rev. Let., 109, 121101
%
\bibitem[\protect\citeauthoryear{Bekenstein \& Oron}{1978}]{bekenstein78} 
Bekenstein J.D., \& Oron E., 1978, PRD, 18, 1809
%
\bibitem[\protect\citeauthoryear{Beskin}{2010}]{beskin10} 
Beskin V.S., 2010, Phys. Uspekhi, 53, 1199
%
\bibitem[\protect\citeauthoryear{Beskin et al.}{1983}]{beskin83} 
Beskin V.S., Gurevich A.V., \& Istomin Y.N., 1983, Sov. Phys. JETP, 58, 235
%
\bibitem[\protect\citeauthoryear{Beskin \& Kusnetsova}{2000}]{beskin00} 
Beskin V.S., \& Kusnetsova I.V., 2000, Nuovo Cimento B, 115, 795
%
\bibitem[\protect\citeauthoryear{Beskin et al.}{1992}]{beskin92} 
Beskin V.S., Istomin Y.N., \& Pariev V.I., 1992, Sov. Astron., 36, 642
%
\bibitem[\protect\citeauthoryear{Beskin \& Rafikov}{2000}]{beskin_rafikov00} 
Beskin V.S., \& Rafikov R.R., 2000, MNRAS, 313, 433
%
\bibitem[\protect\citeauthoryear{Blandford \& Znajek}{1977}]{blandford77}
Blandford R.D., \& Znajek R.L., 1977, MNRAS, 176, 465 
%
\bibitem[\protect\citeauthoryear{Contopoulos et al.}{2013}]{contopoulos13}
Contopoulos I., Kazanas D., \& Papadopoulos D.B., 2013, ApJ, 765, 113
%
\bibitem[\protect\citeauthoryear{Goldreich \& Julian}{1969}]{goldreich69} 
Goldreich P., Julian W.H., 1969, ApJ, 157, 869
%
\bibitem[\protect\citeauthoryear{Goodwin et al.}{2004}]{goodwin04} 
Goodwin S.P., Mestel J., Mestel L., \& Wright G.A.E., 2004, MNRAS, 349, 213
%
\bibitem[\protect\citeauthoryear{Hirotani \& Okamoto}{1998}]{hirotani98} 
Hirotani K., \& Okamoto I., 1998, ApJ, 497, 563
%
\bibitem[\protect\citeauthoryear{Khanna}{1998}]{khanna98} 
Khanna, R., 1998, MNRAS, 294, 673
%
\bibitem[\protect\citeauthoryear{King et al.}{1975}]{king75} 
King A.R., Lasota J.P., \& Kundt W., 1975, Phys. Rev. D, 12, 3037
%
\bibitem[\protect\citeauthoryear{Koide et al.}{2002}]{koide02} 
Koide S., Shibata K., Kudoh T., \& Meier D.L., 2002, Science, 295, 1688
%
\bibitem[\protect\citeauthoryear{Komissarov}{2004a}]{komissarov04} 
Komissarov S.S., 2004a, MNRAS, 350, 427
%
\bibitem[\protect\citeauthoryear{Komissarov}{2004b}]{komissarov04b} 
Komissarov S.S., 2004b, MNRAS, 350, 1431
%
\bibitem[\protect\citeauthoryear{Komissarov}{2005}]{komissarov05} 
Komissarov S.S., 2005, MNRAS, 359, 801
%
\bibitem[\protect\citeauthoryear{Komissarov}{2009}]{komissarov09} 
Komissarov S.S., 2009, J. Korean Phys. Soc., 54, 2503
%
\bibitem[\protect\citeauthoryear{Komissarov \& Barkov}{2009}]{barkov09} 
Komissarov S.S., \& Barkov M.V., 2009, MNRAS, 397, 1153
%
\bibitem[\protect\citeauthoryear{Komissarov \& McKinney}{2007}]{komissarov07} 
Komissarov S.S., \& McKinney J.C., 2007, MNRAS, 377, L49
%
\bibitem[\protect\citeauthoryear{Landau \& Lifshitz}{1975}]{landau75} 
Landau L.D., \& Lifshitz E.M., 1975, The Classical Theory of Fields, 
Course of Theoretical Physics, Vol. 2 (New York: Pergamon)
%
\bibitem[\protect\citeauthoryear{Levinson}{2006}]{levinson06} 
Levinson A., 2006, in Trends in Black Hole Research, edited by Kreitler P.V.
(Nova Science Publishers, New York), 119 (arXiv:astro-ph/0502346)
%
\bibitem[\protect\citeauthoryear{Lovelace}{1976}]{lovelace76} 
Lovelace R.V.E., 1976, Nature, 262, 649
%
\bibitem[\protect\citeauthoryear{McKinney}{2006}]{mckinney06} 
McKinney J.C., 2006, MNRAS, 368, 1561
%
\bibitem[\protect\citeauthoryear{Melatos \& Melrose}{1996}]{melatos96}
Melatos A., \& Melrose D.B., 1996, MNRAS, 279, 1168
%
\bibitem[\protect\citeauthoryear{Menon \& Dermer}{2005}]{menon05}
Menon G., \& Dermer C.D., 2005, ApJ, 635, 1197
%
\bibitem[\protect\citeauthoryear{Menon \& Dermer}{2011}]{menon11}
Menon G., \& Dermer C.D., 2011, MNRAS, 417, 1098
%
\bibitem[\protect\citeauthoryear{Okamoto}{2012}]{okamoto12} 
Okamoto I., 2012, Publ. Astron. Soc. Japan, 64, 50
%
\bibitem[\protect\citeauthoryear{Paczynski}{1990}]{paczynski90} 
Paczynski B., 1990, ApJ, 363, 218
%
\bibitem[\protect\citeauthoryear{Penna et al.}{2013}]{penna13}
Penna R.F., Narayan R., \& Sadowski A., MNRAS, 436, 3741
%
\bibitem[\protect\citeauthoryear{Penrose}{1969}]{penrose69}
Penrose R., 1969, Nuovo. Cim., 1, 252 
%
\bibitem[\protect\citeauthoryear{Punsly}{2008}]{punsly08}
Punsly B., 2008, Black Hole Gravitohydromagnetics, 2nd Edition. Springer, Berlin
%
\bibitem[\protect\citeauthoryear{Punsly \& Coroniti}{1989}]{punsly89} 
Punsly B., \& Coroniti F.V., 1989, Phys. Rev. D, 40, 3834
%
\bibitem[\protect\citeauthoryear{Punsly \& Coroniti}{1990}]{punsly90} 
Punsly B., \& Coroniti F.V., 1990, ApJ, 354, 583
%
\bibitem[\protect\citeauthoryear{Ruiz et al.}{2012}]{ruiz12}
Ruiz M., Palenzuela C., Galeazzi F., \& Bona C., 2012, MNRAS, 423, 1300
%
\bibitem[\protect\citeauthoryear{Takahashi et al.}{1990}]{takahashi90}
Takahashi M., Nitta S., Tatematsu Y., \& Tomimatsu A., 1990, ApJ, 363, 206
%
\bibitem[\protect\citeauthoryear{Tchekhovskoy et al.}{2011}]{tchekho11}
Tchekhovskoy, A., Narayan, R., \& McKinney, J.C., 2011, MNRAS, 418, L79
%
\bibitem[\protect\citeauthoryear{Thorne \& MacDonald}{1982}]{thorne82}
Thorne K.S., MacDonald D.A., 1982, MNRAS, 198, 339
%
\bibitem[\protect\citeauthoryear{Thorne et al.}{1986}]{thorne86} 
Thorne K.S., Price R.H., \& MacDonald D.A., 1986, The Membrane Paradigm. Yale Univ. Press, New Haven, CT
%
\bibitem[\protect\citeauthoryear{Toma \& Takahara}{2012}]{toma12} 
Toma K., \& Takahara F., 2012, ApJ, 754, 148
%
\bibitem[\protect\citeauthoryear{Toma \& Takahara}{2013}]{toma13} 
Toma K., \& Takahara F., 2013, Prog. Theor. Exp. Phys., 2013, 083E02
%
\bibitem[\protect\citeauthoryear{Wald}{1974}]{wald74} 
Wald R.M., 1974, Phys. Rev. D, 10, 1680
%
\end{thebibliography}
\end{document}